%% file: RLonFPGA.tex
\documentclass[aps,prx,amsmath,amssymb,amsfonts,superscriptaddress,notitlepage,reprint,longbibliography,floatfix
]{revtex4-2}

\usepackage{graphicx}
\usepackage{dcolumn}
\usepackage{bm}
\usepackage{amssymb}
\usepackage{amsmath}
\usepackage[svgnames]{xcolor}
\usepackage[colorlinks=true,
            linkcolor=red,
            urlcolor=blue,
            citecolor=DarkGreen]{hyperref}
\usepackage[english]{babel}
\usepackage{braket}
\usepackage{microtype}
\usepackage{bbm}
\usepackage{tikz}
\usepackage{siunitx}
\usepackage{physics}
\usepackage{csquotes}

\usepackage{float}
\makeatletter
\let\newfloat\newfloat@ltx
\makeatother
\usepackage{algorithm}
\usepackage{algpseudocode}

\usepackage[capitalise]{cleveref}

\begin{document}
\title{Realizing a deep reinforcement learning agent \\ discovering real-time feedback control strategies for a quantum system}

\author{Kevin~Reuer}\email{kevin.reuer@phys.ethz.ch}
\affiliation{Department of Physics, ETH Zurich, CH-8093 Zurich, Switzerland}
\author{Jonas~Landgraf}
\author{Thomas~Fösel}
\affiliation{Max Planck Institute for the Science of Light, Staudtstraße 2, 91058 Erlangen, Germany}
\affiliation{Physics Department, University of Erlangen-Nuremberg, Staudtstraße 5, 91058 Erlangen, Germany}
\author{James~O'Sullivan}
\author{Liberto~Beltrán}
\author{Abdulkadir~Akin}
\author{Graham~J.~Norris}
\author{Ants~Remm}
\author{Michael~Kerschbaum}
\author{Jean-Claude~Besse}
\affiliation{Department of Physics, ETH Zurich, CH-8093 Zurich, Switzerland}

\author{Florian~Marquardt}
\affiliation{Max Planck Institute for the Science of Light, Staudtstraße 2, 91058 Erlangen, Germany}
\affiliation{Physics Department, University of Erlangen-Nuremberg, Staudtstraße 5, 91058 Erlangen, Germany}

\author{Andreas~Wallraff}
\affiliation{Department of Physics, ETH Zurich, CH-8093 Zurich, Switzerland}
\affiliation{Quantum Center, ETH Zurich, CH-8093 Zurich, Switzerland}

\author{Christopher~Eichler}\email{christopher.eichler@fau.de}
\altaffiliation[Current address: ]{Physics Department, University of Erlangen-Nuremberg, Staudtstraße 5, 91058 Erlangen, Germany}
\affiliation{Department of Physics, ETH Zurich, CH-8093 Zurich, Switzerland}

\date{\today}

\begin{abstract}
To realize the full potential of quantum technologies, finding good strategies to control quantum information processing devices in real time becomes increasingly important.
Usually these strategies require a precise understanding of the device itself, which is generally not available.
Model-free reinforcement learning circumvents this need by discovering control strategies from scratch without relying on an accurate description of the quantum system. Furthermore, important tasks like state preparation, gate teleportation and error correction need feedback at time scales much shorter than the coherence time, which for superconducting circuits is in the microsecond range. Developing and training a deep reinforcement learning agent able to operate in this real-time feedback regime has been an open challenge. Here, we have implemented such an agent in the form of a latency-optimized deep neural network on a field-programmable gate array (FPGA). We demonstrate its use to efficiently initialize a superconducting qubit into a target state.
To train the agent, we use model-free reinforcement learning that is based solely on measurement data.
We study the agent's performance for strong and weak measurements, and for three-level readout, and compare with simple strategies based on thresholding.
This demonstration motivates further research towards adoption of reinforcement learning for real-time feedback control of quantum devices and more generally any physical system requiring learnable low-latency feedback control.

\end{abstract}

\maketitle

Future quantum information processing devices will rely on the ability to continuously monitor their state via quantum measurements and to act back on them, on timescales much shorter than the coherence time, conditioned on prior observations. Such real-time feedback control of quantum systems, which offers applications e.g. in qubit initialization \cite{Riste2012b,Salathe2018,Negnevitsky2018}, gate teleportation \cite{Steffen2013,Chou2018} and quantum error correction \cite{Ofek2016,Andersen2019,RyanAnderson2021}, typically relies on an accurate model of the underlying system dynamics. 
With the increasing number of constituent elements in quantum processors such accurate models are generally not available. Model-free reinforcement learning \cite{sutton2018reinforcement} promises to overcome such limitations by learning feedback-control strategies without prior knowledge of the quantum system.

Reinforcement learning, a subfield of machine learning, has had outstanding success in tasks ranging from board games \cite{silver2016mastering} to robotics \cite{kober2013reinforcement}.
Reinforcement learning, however, has only very recently been started to be applied to complex physical systems, with training performed either on simulations \cite{Bellemare2020,praeger2021playing,Guo2021,Ai2022,kuprikov2022deep,degrave2022magnetic,Peng2022} or directly in experiments \cite{tunnermann2019deep,kain2020sample,hirlaender2020model,yan2021low,MuinosLandin2021,Baum2021a,Nguyen2021c,li2022deep}, for example in laser \cite{tunnermann2019deep,yan2021low,li2022deep}, particle \cite{kain2020sample,hirlaender2020model}, soft-matter \cite{MuinosLandin2021} and quantum physics \cite{Baum2021a,Nguyen2021c}.
Specifically in the quantum domain, during the past few years, a number of theoretical works have pointed out the great promises of reinforcement learning for tasks covering state preparation \cite{chen2013fidelity,Bukov2018,borah2021measurement,sivak2021model,porotti2022deep}, gate design \cite{niu2019universal}, error correction \cite{Fosel2018,nautrup2019optimizing,sweke2020reinforcement} and circuit optimization/compilation \cite{zhang2020topological,fosel2021quantum}, making it an important part of the machine learning toolbox for quantum technologies \cite{Carleo2019,Dawid2022,Krenn2022}.
In first applications to quantum systems, reinforcement learning was experimentally deployed, but training was mostly performed based on simulations, specifically to optimize pulse sequences for atoms and spins \cite{Guo2021,Ai2022,Peng2022}. Beyond that, there are two pioneering works demonstrating the training directly on experiments \cite{Nguyen2021c,Baum2021a} which was used to optimize pulses for quantum gates \cite{Baum2021a} and to accelerate the tune-up of quantum dot devices \cite{Nguyen2021c}. However, in none of these experiments \cite{Guo2021,Peng2022,Ai2022,Nguyen2021c,Baum2021a} real-time quantum feedback was required. 

Here, we realize a reinforcement learning agent which interacts with a quantum system on a sub-microsecond timescale. This rapid response time enables the agent's use for real-time quantum feedback control.
We implement the agent using a novel low-latency neural network architecture, which processes data concurrently to data acquisition, on a field-programmable gate array (FPGA).
As a proof of concept, we train the agent using model-free reinforcement learning to initialize a superconducting qubit into its ground state without relying on a prior model of the quantum system. The training is performed directly on the experiment, i.e.~by acquiring experimental data with updated network parameters in every training step.
In repeated cycles, the trained agent acquires measurement data, processes it and applies pre-calibrated pulses to the qubit conditioned on the measurement outcome until the agent terminates the initialization process.  
We study the evolution of the agent's performance during training and demonstrate convergence in less than three minutes wall clock time and based on less than 30,000 episodes of training data. Furthermore, we explore the agent's strategies in more complex scenarios, i.e. when performing weak measurements or when resetting a qutrit.

\section{Reinforcement learning for a quantum system}
\begin{figure}[!t]
\includegraphics[width=\columnwidth]{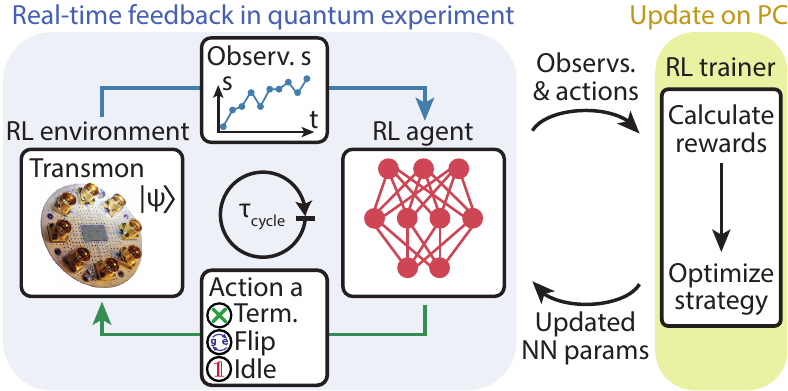}
\caption{\label{RL1} Concept of the experiment. A reinforcement learning (RL) agent, realized as a neural network (red) on a field-programmable gate array (FPGA), receives observations $\vectorbold{s}$ (blue trace) from a quantum system, which constitutes the reinforcement learning (RL) environment. Here, the quantum system is realized as a transmon qubit coupled to a readout resonator fabricated on a chip (see photograph). The agent processes observations on sub-microsecond timescales to decide in real time on the next action $a$ applied to the quantum system. The update of the agent's parameters is performed by processing experimentally obtained batches of observations and actions on a PC.
}
\end{figure}

In model-free reinforcement learning, an agent interacts with the world around it, the so-called reinforcement learning environment (see Fig.~\ref{RL1}). In repeated cycles, the agent receives observations ${\mathbf s}$ from the environment and selects actions $a$ according to the respective observation ${\mathbf s}$ and its policy $\pi$. In the important class of policy-gradient methods \cite{sutton2018reinforcement}, this policy is realized as a conditional probability distribution $\pi_{\vectorbold{\theta}}(a|{\mathbf s})$, which can be modelled as a neural network with parameters $\vectorbold{\theta}$. To each sequence of observation-action pairs, called an episode, one assigns a cumulative reward $R$. The goal of reinforcement learning is to maximize the reward $\bar{R}$ averaged over multiple episodes, by updating the parameters $\vectorbold{\theta}$ e.g. via gradient ascent $\Delta \theta \sim \nabla_{\vectorbold{\theta}} \bar{R}$ \cite{sutton2018reinforcement}. Such a policy-gradient procedure is able to discover an optimal policy even without access to an explicit model of the reinforcement learning environment's dynamics.

In the present work, our goal is to use reinforcement learning to learn strategies for real-time control of quantum systems. Here, observations are obtained via quantum measurements, actions are realized as unitary gate operations, and the reward is measured in terms of the speed and fidelity of initializing the quantum system into a target state, see schematic in Fig.~\ref{RL1}. In our experiment the quantum system is realized as a transmon qubit with ground $\ket{g}$, excited $\ket{e}$, and second excited state $\ket{f}$ dispersively coupled to a superconducting resonator (see App.~\ref{setup} for details).
We probe the qubit with a microwave field, which scatters off the resonator and gets amplified and digitized to result in an observation vector $\vectorbold{s}=(\vectorbold{I},\vectorbold{Q})$, where $\vectorbold{I}$ and $\vectorbold{Q}$ are time traces of the two quadrature components of the digitized signal \cite{Blais2004,Wallraff2005,Gambetta2007} (see App.~\ref{RO} for details and App.~\ref{Disc} for averaged time traces).
Depending on $\vectorbold{s}$ the agent selects, according to its policy $\pi$, one of several discrete actions in real time. In the simplest case, it either \emph{idles} until the next measurement cycle, it performs a \emph{bit-flip} as a unitary swap between $\ket{g}$ and $\ket{e}$ or it \emph{terminates} the initialization process.

To train the agent, we transfer batches of episodes to a personal computer (PC) serving as a reinforcement learning trainer. The reinforcement learning trainer computes the associated reward for each episode and updates the agent's policy accordingly (see App.~\ref{Train} for details), before sending back the updated network parameters $\theta$ to the FPGA.

\section{Implementation of the neural-network-based real-time agent}
\begin{figure}[!t]
\includegraphics[width=\columnwidth]{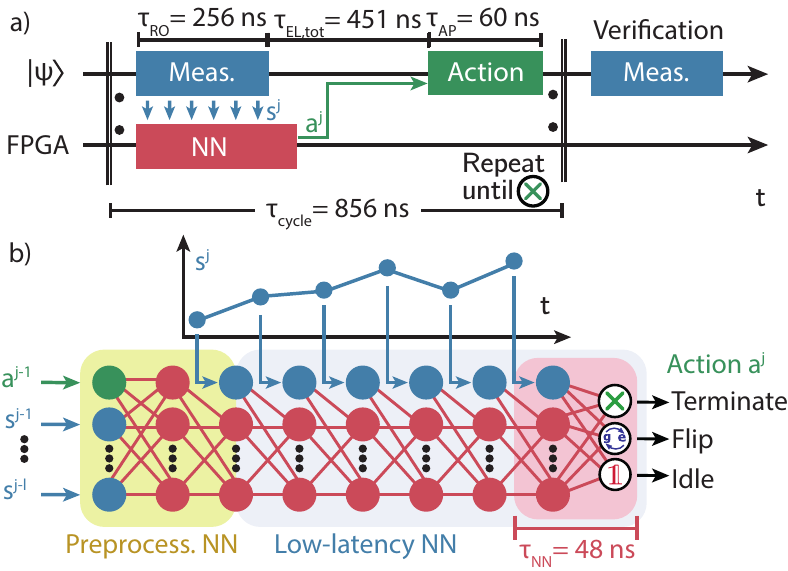}
\caption{\label{FPGA} Schematic of neural-network-based real-time feedback control. (a) Timing diagram of a reinforcement learning episode.
In each cycle $j$, the observation $\vectorbold{s}^j$ resulting from a measurement (blue) is continuously fed into a neural network (red) which decides on the next action $a^j$ (green). Once the agent terminates after several cycles, a verification measurement is performed.
(b) Schematic of the neural network implemented on an FPGA. The neural network consists of fully-connected (red lines) layers of feed-forward neurons (red dots) and input neurons (blue dots for observations, green dots for actions).
The first layers form the preprocessing network (yellow background).
During the evaluation of the low-latency network (blue background), new data points from the signal trace $\vectorbold{s}^j$ are fed into the network as they become available.
The network outputs the action probabilities for the three actions. Only the execution of the last layer (red background) contributes to the overall latency.
}
\end{figure}

We implemented this scheme in an experimental setup, in which the agent, for each episode, is able to perform multiple measurement cycles $j$, in each of which it receives a qubit-state-dependent observation $\vectorbold{s}^j$ and selects an action $a^j$, until it terminates the episode, see Fig.~\ref{FPGA}(a). If the agent selects the flip action, a $\pi$-pulse is applied after a total latency of $\tau_{\text{EL,tot}}=451$~ns, dominated by analog-to-digital and digital-to-analog converter delays. The agent's neural network contributes only $\tau_{\text{NN}}=48$~ns to the total latency as it is evaluated mostly during qubit readout and signal propagation (see App.~\ref{RO} for detailed discussion of the latency). To provide the agent with a memory about past cycles we feed downsampled observations $(\vectorbold{s}^{j-1}, ..., \vectorbold{s}^{j-l})$ and actions $(a^{j-1}, ..., a^{j-l})$ from up to $l=2$ previous cycles into the neural network. To characterize the performance of the agent, we perform a verification measurement $\vectorbold{s}^{\text{ver}}$ after termination.

Any neural-network agent used for real-time system control greatly benefits from short latencies in the signal processing. For our FPGA implementation we therefore introduce a novel network architecture, which aims to keep latencies at a minimum, see Fig.~\ref{FPGA}(b). First, we implement the agent as a feedforward neural network \cite{Goodfellow2016} on the FPGA, rather than a more resource-demanding recurrent neural network, like a long short-term memory network (LSTM) \cite{Hochreiter1997}. Second, we process information from previous cycles in a two-layer \emph{pre-processing network} before the start of the current cycle, thus not contributing to the latency. Third, and most importantly, we implement a novel \emph{low-latency network} architecture, in which new measurement data is processed as soon as it becomes available. More specifically, we sequentially feed elements $I_k^j$, $Q_k^j$ of the digitized time trace $\vectorbold{s}^j=(\vectorbold{I}^j,\vectorbold{Q}^j)$ into each layer of the neural network concurrent with its evaluation, see Fig.~\ref{FPGA}(b) and App.~\ref{sec:min_latency_network}. As a result, only the execution of the last layer contributes to the total latency while all other layers are evaluated in parallel with the data acquisition. For the experiments presented in the following, we use a network with 7 hidden layers and 12 neurons per layer. The output layer has only three neurons, corresponding to the three actions. However, as the exact neural network structure may in general depend on the properties of the specific quantum system and the agent's task, the width and depth of the neural network are adjustable parameters in our FPGA design. We have also explored the use of the same type of neural network for quantum state discrimination, in a supervised-learning setting (see App.~\ref{Disc})  .                                                 

\begin{figure*}[!t]
\includegraphics[width=\textwidth]{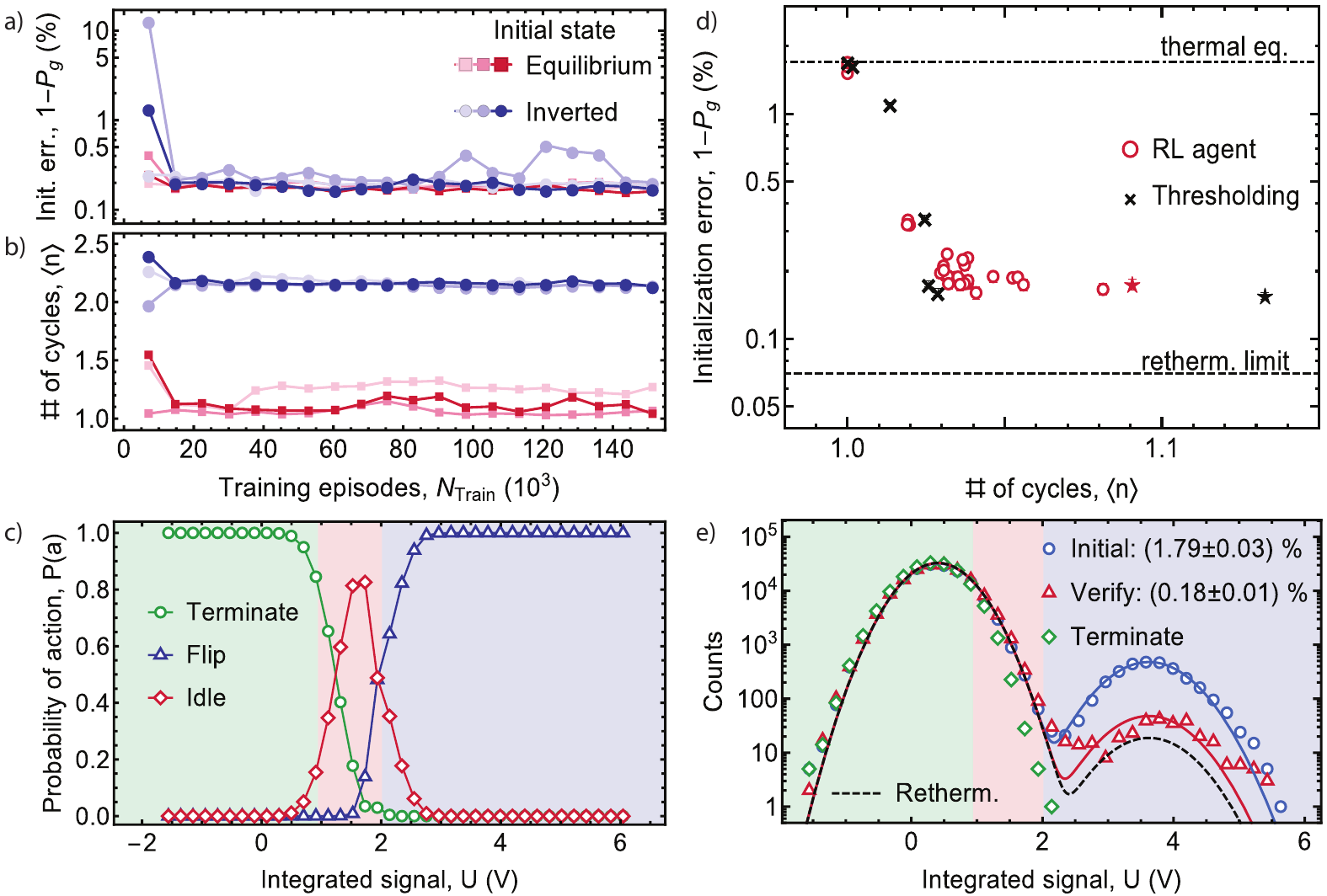}
\caption{\label{RL}
Experimental data for reinforcement learning with a network-based real-time agent. (a) Initialization error $1-P_g$ and (b) average number of cycles $\langle n \rangle$ until termination vs number of training episodes $N_{\text{train}}$, when preparing an equilibrium state (red squares) and when inverting the population with a $\pi$-pulse (dark blue circles) for three independent training runs (solid and transparent points). Each datapoint is obtained from an independent validation data set with $\sim 180,000$ episodes. (c) Probability of choosing an action $P(a)$ vs the integrated measurement signal $U$. Actions chosen by the threshold-based strategy are shown as background colors (also for (e)). (d) Initialization error $1-P_g$ vs average number of cycles $\langle n \rangle$ until termination for an equilibrium state for the reinforcement learning agent (red circles) and the threshold-based strategy (black crosses). Stars indicate the strategies used for the experiments in (c) and (e). (e) Histogram of $U$ in equilibrium (blue circles), for the measurement in which the agent terminates (green diamonds) and for the verification measurement (red triangles). Lines are bimodal Gaussian fits, from which we extract ground state populations as shown in the inset. The dashed black line indicates the rethermalization limit (see main text).}
\end{figure*}

\section{Training the agent with experimental data}
To train the agent, we experimentally acquire 1000 observation-action pairs $(\vectorbold{s},a)$ with the FPGA, before transferring them to the reinforcement learning trainer on the PC, see Fig.~\ref{RL1}. The reinforcement learning trainer then updates the parameters $\vectorbold{\theta}$ of the agent's policy $\pi_{\vectorbold{\theta}}(a|\vectorbold{s})$ using a state-of-the-art algorithm (proximal policy optimization, PPO) \cite{schulman_proximal_2017, hill_stable_2018}, with the goal to maximize the cumulative reward $R = U_{\text{ver}}/\Delta U - n \lambda $ (see App.~\ref{Train} for details). Here, the integrated observation in the final verification measurement $U_{\text{ver}}=\vectorbold{w_s} \vectorbold{s}^{\text{ver}}$ serves as an indicator for the ground-state population, with a normalization factor $\Delta U=\vectorbold{w_s} \left(\langle \vectorbold{s}_g \rangle - \langle \vectorbold{s}_e \rangle \right) $ setting the scale, and the second term penalizes each cycle with a constant amount $\lambda$. Thus, $\lambda$ controls the trade-off between short episode length and high initialization fidelity. The updated parameters $\vectorbold{\theta}$ are then transferred back to the FPGA, and we repeat this procedure until the cumulative reward $R$ is maximized. 

We first train the reinforcement learning agent to initialize the qubit using fast, high-fidelity readout. In this regime, an initialization strategy based on weighted integration and thresholding is close-to-optimal, and we can thus easily verify and benchmark the strategies discovered by the reinforcement learning agent.
To study the agent's learning process, we monitor the average initialization error $1-P_g$, inferred from a fit to the measured distribution of $U_{\text{ver}}$ (see App.~\ref{RO} for details), and the average number of cycles $\langle n \rangle$ until termination, see Fig~\ref{RL}(a) and (b). While the initial random policy results in only $P_g\sim 50\%$, the agent quickly learns how to initialize the qubit for both prepared initial states, which we choose to be the equilibrium state (red) and its counterpart with the population inverted by a $\pi$-pulse (dark blue). The initialization error $1-P_g$ has already converged to about $0.2\,\%$ after training with about 30,000 episodes, which includes 100 parameter updates by the reinforcement learning trainer on the PC, and takes only three minutes wall clock time. The short training duration, limited mainly by data transfer between the PC and FPGA, enables frequent readjustment of the neural network parameters and thus allows to account for drifts in experimental parameters. The average number of cycles $\langle n \rangle$ converges to about 1.1 for the initial equilibrium, and to about 2.2 for the inverted equilibrium state, indicating that for strong measurements the agent only needs an additional cycle to terminate for about $10$~$\%$, or respectively $20$~$\%$, of the episodes. 

\section{Policy and performance for strong measurements}
After the training has been completed, we analyze the policy. Instead of directly investigating the (high-dimensional) dependence of the agent's policy $\pi(a|\vectorbold{s})$ on the full time trace $\vectorbold{s}$, we simply extract the probabilities $P(a)$ for the agent to select the possible actions depending on the integrated signal $U$, see Fig.~\ref{RL}(c). We compare the agent's selection of actions to a simple strategy in which the process is terminated (green background) for $U$ values below an acceptance threshold and a flip (blue background) is applied for $U$ larger than a state discrimination threshold.
We observe that for $U$ far below the acceptance threshold the agent nearly always terminates, while the agent predominantly selects the flip action for $U$ far above the state discrimination threshold. This is expected as, in both cases, the agent has high certainty about the qubit state. Between the two thresholds where uncertainty is large, the agent is more likely to idle. The transitions of the individual probabilities are smooth. This is not due to some deliberate randomization of action choices, but rather a sign that the agent's policy depends on additional information beyond the integrated signal $U$ shown here: the agent has access to the full measured time trace.

To evaluate the agent's performance we analyze the tradeoff between initialization error $1-P_g$ and average cycle number $\langle n \rangle$ as a function of the control parameter $\lambda$. As expected, we find that an increase in $\langle n \rangle$, controlled by lowering $\lambda$, results in a gain of initialization fidelity until $1-P_g$ converges to about $0.18\%$ (for $\langle n \rangle \geq 1.1$ cycles), see Fig~\ref{RL}(d). 
We attribute the remaining infidelity mostly to rethermalization of the qubit between the termination and the verification cycle, and, possibly, state mixing during the final verification readout. In our experiment, this rethermalization rate is $N_{\text{eq}}/T_1 \approx 1$~kHz with $N_{\text{eq}}=1.4\,\%$, contributing $\sim 0.07\,\%$ to the infidelity.
As anticipated, the agent's performance matches the performance of simple, close-to-optimal, thresholding strategies, where we vary the acceptance threshold to control the average cycle number $\langle n \rangle$ (black crosses). This indicates that the strategies discovered by the agent are also close-to-optimal.
We also deduce that the agent's performance is limited mostly by rethermalization by analyzing the bimodal Gaussian distribution of the integrated qubit readout signal $U_{\text{ver}}$ in the verification measurement (red triangles in Fig.~\ref{RL}(e)). While the integrated signal in the termination cycle (green diamonds) has only very few counts above the state discrimination threshold, the number of such instances rises to about $0.18\%$ in the verification measurement, indicating transitions into the excited state occuring between the two cycles. Compared to the equilibrium state (blue circles) the excited state fraction is reduced by about a factor 10 by using the reinforcement learning initialization scheme.

\section{Weak measurements and qutrit readout}
\begin{figure*}[!t]
\includegraphics[width=\textwidth]{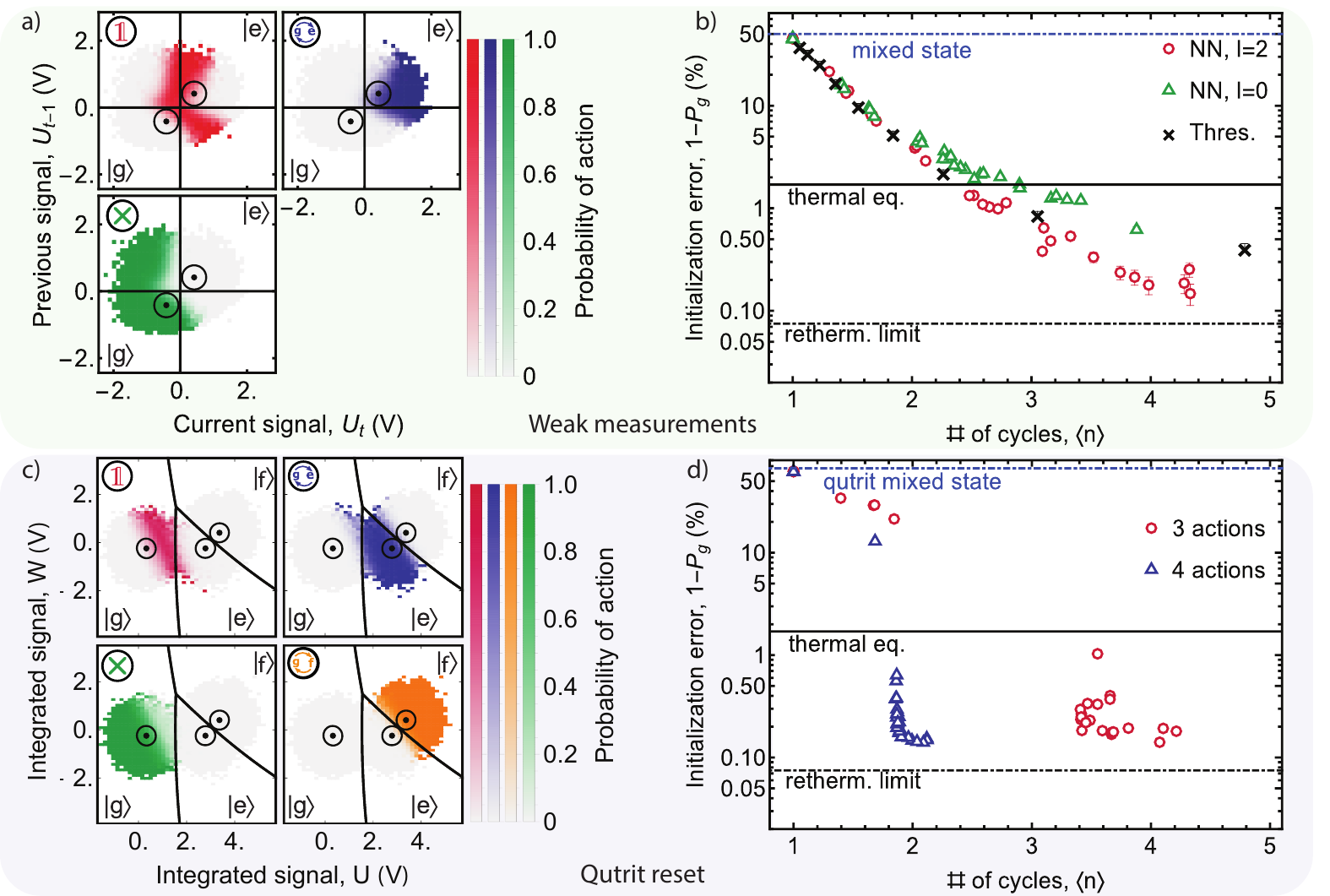}
\caption{\label{RL2}
Reinforcement learning results for weak measurements and three-level systems. (a) Probability $P(a)$ of choosing the action indicated in the top left corner vs the signal of the current $U_t$ and the previous $U_{t-1}$ cycle for $l=2$. The radii of the black circles indicate the standard deviation around the means (black dots) of the fitted bi-modal Gaussian distribution. Black lines are the state discrimination thresholds (normalized to 0, see App.~\ref{RO}). $P(a)$ is shown for each bin with at least a single count. Empty bins are colored white (also for (d)). (b) Initialization error $1-P_g$ vs $\langle n \rangle$ for weak measurements for an initially mixed state for $l=2$ (red circles), $l=0$ (green triangles) and a thresholding strategy (black crosses). Performance for $l=1$ (not shown) is similar to $l=2$. (c) Probability $P(a)$ of choosing the action indicated in the top left corner vs $U$ and $W$. Black circles indicate the standard deviation ellipse around the means (black dots) of the fitted tri-modal Gaussian distribution. Black lines are the state discrimination thresholds (see App.~\ref{RO}). (d) Initialization error $1-P_g$ for a completely mixed qutrit state vs $\langle n \rangle$ when the agent can select to \emph{idle}, \emph{flip} and \emph{terminate} (red circles), and when the agent can in addition perform a $gf$-flip (blue triangles).}
\end{figure*}

The observations until this point demonstrate that our real-time agent performs well and trains reliably on experimentally obtained rewards. Next, we discuss regimes where good initialization strategies are more complex. As a first example, we investigate the agent's strategy and performance when only weakly measuring the qubit. We reduce the power of the readout tone, while keeping its duration and frequency unchanged, such that bimodal Gaussian distributions of a prepared ground and excited state overlap by 25 \% (see App.~\ref{RO}).
In this case, we find that the agent profits from memory, if it is permitted access to information from $l$ previous cycles. In that case, its strategy not only depends on the current signal $U_t$, but also on the signal $U_{t-1}$ from the previous cycle, see Fig.~\ref{RL2}(a).
Whenever the current measurement hints at the same state as the previous measurement (upper right and lower left in each panel) the agent gains certainty about the state and thus becomes more likely to terminate the process (green region in lower left corner) or swap the $\ket{g}$ and the $\ket{e}$ state (blue region in upper right corner).
As for strong measurements, we find a trade-off between $\langle n \rangle$ and $1-P_g$ when varying $\lambda$, see Fig.~\ref{RL2}(b). Importantly, we observe that agents making use of memory ($l=2$, red circles) require fewer rounds $\langle n \rangle$ to reach a certain initialization error than agents without memory ($l=0$, green triangles) or a thresholding strategy (black crosses).
This indicates that the observed dependence of the strategy on $U_{t-1}$ does result in a performance improvement; the reinforcement learning agent can exploit the possibility of measuring multiple times, in contrast to the simple thresholding strategy.

In addition, we have studied the performance of the agent when also considering the second excited state $\ket{f}$, which we have neglected so far. The $\ket{f}$ state is populated with a certain probability due to undesired leakage out of the computional states $\ket{g}$ and $\ket{e}$ during single-qubit, two-qubit and readout operations \cite{Magnard2018}. Thus, schemes which also reset $\ket{f}$ into $\ket{g}$ are required. For this purpose, we enable the agent to also swap $\ket{f}$ and $\ket{g}$ states by adding a fourth action, and train the agent on a qutrit mixed state with one third $\ket{g}$, $\ket{e}$ and $\ket{f}$ population, prepared by idling the qubit, swapping the qubits $\ket{g}$ and $\ket{e}$ state, and swapping the qubits $\ket{g}$ and $\ket{f}$ state, with probability $1/3$ respectively.
For this qutrit system, state assignment typically processes two different projections of the measurement trace $U=\vectorbold{w_U} \vectorbold{s}^{\text{ver}}$ and $W=\vectorbold{w_W} \vectorbold{s}^{\text{ver}}$, where $\vectorbold{w_U}$ and $\vectorbold{w_W}$ form an orthonormal set of weights. Here, we use $U$ and $W$ to visualize the agent's strategy.
We find that, if the qubit would be classified to be in the $\ket{f}$ state, the agent is most likely to swap $\ket{f}$ and $\ket{g}$ (orange), as expected, see Fig.~\ref{RL2}(d). Similarly, if the qubit is classified to be in $\ket{g}$, the agent most likely terminates, while the agent swaps $\ket{e}$ and $\ket{g}$, if it is classified to be in $\ket{e}$. Around the state discrimination threshold between $\ket{g}$ and $\ket{e}$, however, the agent mostly idles, as the qubit state is uncertain. Interestingly, along the state discrimination threshold between $\ket{e}$ and $\ket{f}$, the agent almost never idles, as it is more advantageous either to apply a flip from $\ket{e}$ to $\ket{g}$ or a flip from $\ket{f}$ to $\ket{g}$. After the operation, the qubit will be in the ground state if, by chance, the correct action was applied, while it would never be in the ground state if the agent chose to idle.

To study the performance of the reinforcement learning agent, we trained the agent on a qutrit mixed state. We find that an agent that can swap $\ket{f}$ to $\ket{g}$, in addition to the other actions, efficiently resets the transmon from a qutrit mixed state with an initialization error $1-P_g \approx 0.2 \%$ for $\langle n \rangle \approx 2$ (blue squares), see Fig.~\ref{RL2}(d).
In contrast, an agent which cannot access the $gf$-flip action needs significantly more rounds till termination to reach a similar initialization error, as the agent needs to rely on decay from the $\ket{f}$ level, which in our setup had a lifetime of $T_1^{(f)}=6$~$\SIUnitSymbolMicro$s.

These examples demonstrate the versatility of the reinforcement learning approach to discovering state initialization strategies under a variety of circumstances.

\section{Conclusion}
In conclusion, we have implemented a real-time neural-network agent with a sub-microsecond latency enabled by a network design which accepts data concurrently with its evaluation. This is about 100 times faster than in the fusion control experiments of Ref.~\cite{degrave2022magnetic}, which is, to our knowledge, the fastest reinforcement learning agent deployed in a physics experiment so far. The need for such optimized real-time control will increase due to the ever more stringent requirements on the fidelities of quantum processes as quantum devices grow in size and complexity. We have successfully trained the agent using reinforcement learning in a quantum experiment and demonstrated its ability to adapt its strategy in different scenarios, including those for which memory is beneficial. Our experiments are a first example of reinforcement learning of real-time feedback control on a quantum platform. Applying this method to larger systems will enable the discovery of new strategies for tasks like quantum error correction \cite{Fosel2018,nautrup2019optimizing,sweke2020reinforcement} and many-body feedback cooling \cite{Bukov2018,borah2021measurement,sivak2021model,porotti2022deep}.

\section*{Data availability statement}
The data produced in this work is available from the corresponding authors upon reasonable request.

\section*{Acknowledgments}
This work was supported by the Swiss National Science Foundation (SNSF) through the project ``Quantum Photonics with Microwaves in Superconducting Circuits'', by the European Research Council (ERC) through the project ``Superconducting Quantum Networks'' (SuperQuNet), by the National Centre of Competence in Research ``Quantum Science and Technology'' (NCCR QSIT), a research instrument of the Swiss National Science Foundation (SNSF), by ETH Zurich, the Munich Quantum Valley, which is supported by the Bavarian state government with funds from the Hightech Agenda Bayern Plus, and by the Max Planck Society.

\section*{Author Contributions}
K.R. and J.O. prepared and calibrated the experimental setup. K.R., L.B., and A.A. implemented the neural network on the FPGA. F.M. and C.E. conceived the idea for the experiment. J.L., T.F., and F.M. simulated the system and the network and determined the network structure, training algorithm and reward function, with input from K.R. and C.E.. G.J.N., M.K., A.R., and J.-C.B. fabricated the device. K.R. and J.O. carried out the experiments and analyzed the data, with support from J.L.. K.R., J.L., F.M., and C.E. wrote the manuscript with input from all co-authors. F.M., A.W., and C.E. supervised the work.

\section*{Competing interests}
The authors declare no competing interests.

\begin{appendix}
\input{supplement}

\end{appendix}
\bibliography{bibi.bib}
\end{document}

%% file: supplement.tex
\section{Experimental setup and device calibration} \label{setup}
\begin{figure}[!t]
\includegraphics[width=\columnwidth]{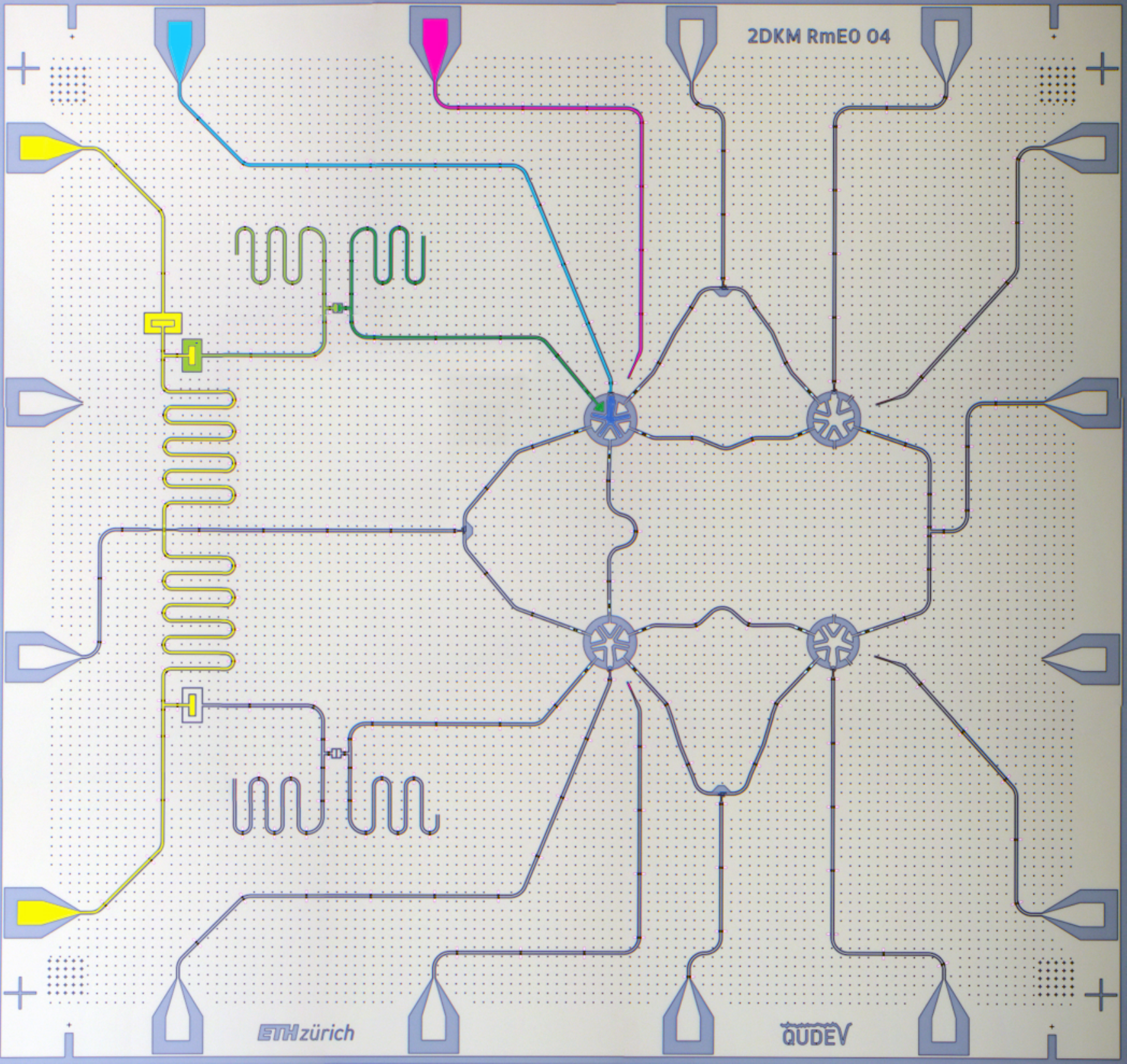}
\caption{\label{chip}
False color optical micrograph of the sample of the used transmon qubit (blue). Depicted are the readout resonator (green) coupled to the feed line (yellow) via a Purcell filter (light green). A flux line (cyan) and a charge line (pink) couple to the qubit. Uncolored parts of the chip are not used.}
\end{figure}

\begin{figure*}[!t]
\includegraphics[width=\textwidth]{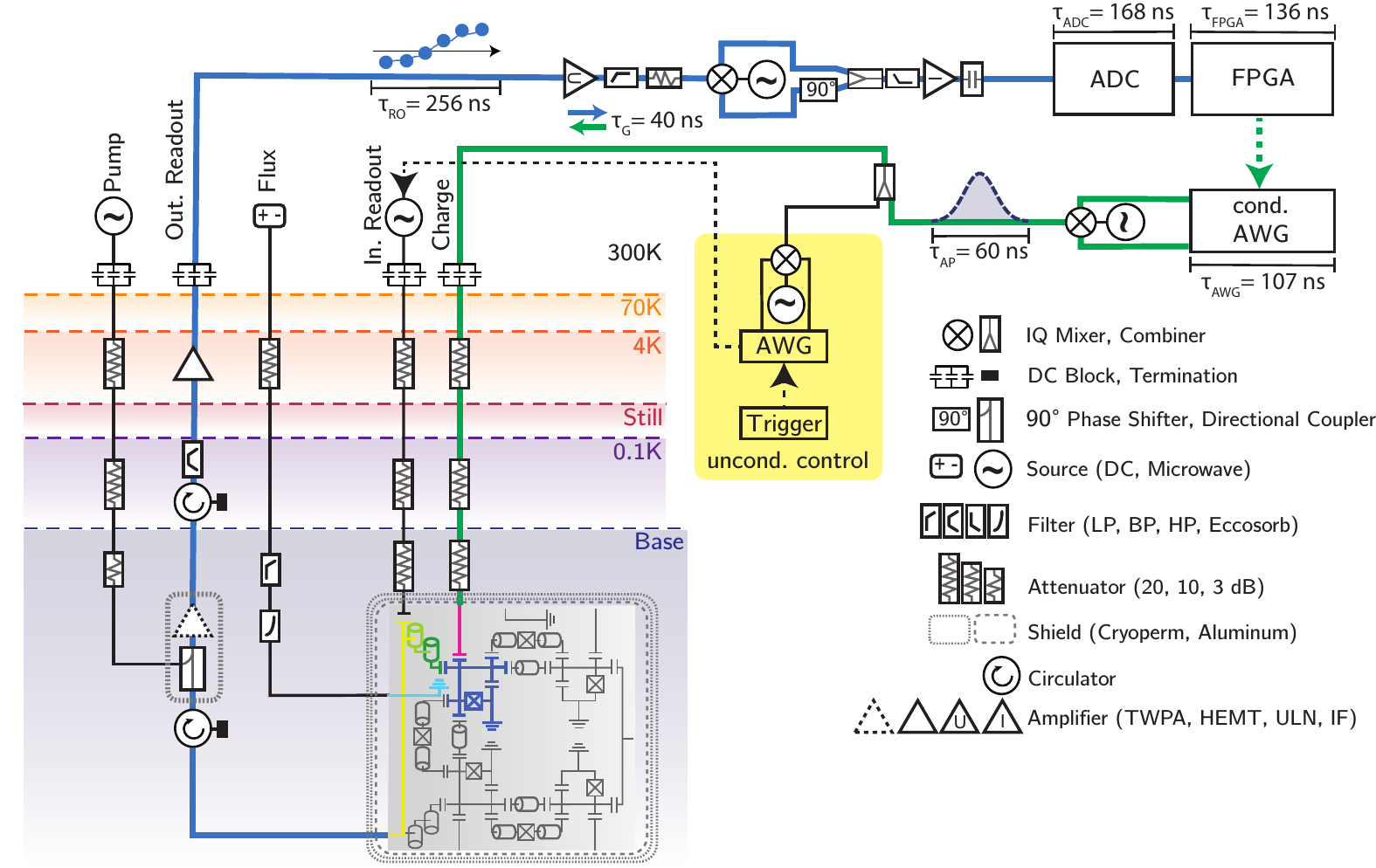}
\caption{\label{figsetup}
Experimental setup for operating the transmon qubit. For details, see text.}
\end{figure*}

For the experiments, we use a transmon qubit coupled to a readout resonator on the chip shown in Fig.~\ref{chip}. The chip is mounted on the base temperature stage (20~mK) of a dilution refrigerator and housed inside three magnetic shields, two made from cryoperm, one from aluminum, see sketch of the experimental setup in Fig.~\ref{figsetup}. We apply microwave pulses to the chip via charge lines with 20 dB attenuation each on the 4~K, 100~mK and base temperature stage for signal conditioning \cite{Krinner2019}. To adjust the qubit frequency, we change the magnetic flux in its superconducting quantum interference device (SQUID) loop by generating currents in an inductively coupled flux line. 

\begin{table}[!b]
\centering
 \begin{tabular}{l  r}
 & \\ 
 \hline
 $g$-$e$ frequency, $\omega_{ge}/2\pi$ [GHz] & 6.524  \\
 $e$-$f$ frequency, $\omega_{ef}/2\pi$ [GHz] & 6.316  \\
 anharmonicity, $\alpha/2\pi$ [MHz] & -209 \\
 lifetime of $\ket{e}$, $T_1^{(e)}$ [$\SIUnitSymbolMicro$s]& 13  \\
 lifetime of $\ket{f}$, $T_1^{(f)}$ [$\SIUnitSymbolMicro$s]& 6  \\
 dephasing time of $\ket{e}$, $T_2^{\star(e)}$ [$\SIUnitSymbolMicro$s] & 2 \\
 dephasing time of $\ket{f}$, $T_2^{\star(f)}$ [$\SIUnitSymbolMicro$s] & 3  \\
 equilibrium excited state population, $P_{\text{therm}}$ [\%] & 1.4 \\
 readout frequency, $\omega_{\text{ro}}/2\pi$ [GHz] & 7.259  \\
 dispersive shift, $\chi/2 \pi$ [MHz] & 10.4 \\
 \hline
 \end{tabular}
 \caption{\label{table:paramp}Measured device parameters.}
\end{table}

To readout the qubit, we generate a 256-ns-long microwave pulse at the readout frequency $\omega_{\text{ro}}$ with a microwave generator (MWG) and apply it to the readout resonator combined with its Purcell filter through the input line. The response of the resonator is then amplified by a traveling wave parametric amplifier (TWPA) with 20~dB gain, a high-electron mobility transistor (HEMT) and a room-temperatur amplifier (blue line in Fig.~\ref{figsetup}). We down-convert the readout signal to 250~MHz, using a local oscillator and an $IQ$ mixer. For image rejection, we re-combine the $I$ and $Q$ channels of the $IQ$ mixers using an $IQ$ combiner, which adds a 90° phase shift to the $Q$ channel. After further filtering and amplification, we digitize the signal with an analog-to-digital converter (ADC) and forward it to an FPGA. In the reinforcement learning approach for initializing the qubit (see main text), the agent on the FPGA then selects an action, and if \emph{flip} is chosen, triggers an arbitrary waveform generator (AWG) (dashed green line). The AWG then plays a pre-programmed derivative removal by adiabatic gate (DRAG) pulse \cite{Motzoi2009}, which is up-converted to the qubit frequency using a local oscillator and an $IQ$ mixer. We then combine this conditional pulse with a periodically-triggered pulse channel used for preparation pulses and apply it to the qubit via a charge line (green line).

Using this setup, we achieve a feedback latency, defined as the time between the end of the readout pulse and the start of the conditional $\pi$-pulse at the qubit, of $\tau_{\text{EL,tot}}\approx 451$~ns. Main contributors to the latency are the ADC ($\tau_{\text{ADC}}=160$~ns, including the delay to generate a trigger for the AWG), the AWG ($\tau_{\text{AWG}} = 107$~ns) and the FPGA ($\tau_{\text{FPGA}}=144$~ns). In addition, there is $\tau_{\text{G}} \approx 40$~ns delay due the signal propagation from the room-temperature electronics to the sample and back through in total about 6 meters of coaxial cable. The latency of the neural network $\tau_{\text{NN}}=48$~ns is included in the FPGA latency $\tau_{\text{FPGA}}$; inefficient signal pre-processing is responsible for 88~ns.

Considering the readout pulse duration $\tau_{\text{RO}}=256$~ns and the duration of the $\pi-$pulse flipping the qubit $\tau_{\text{AP}}=60$~ns, we obtain a minimum cycle duration $\tau_{\text{cycle,min}}=\tau_{\text{RO}}+ \tau_{\text{AP}}+ \tau_{\text{EL,tot}} \approx 767$~ns. When including a \emph{gf-flip}, implemented as a $\pi$-pulse on the $\ket{f}$-$\ket{e}$ manifold, followed by a $\pi$-pulse on the $\ket{e}$-$\ket{g}$ manifold ($\tau_{\text{AP}}=112$~ns), we find $\tau_{\text{cycle,min}}=819$~ns. For the presented experiments, we have chosen a cycle time of 856~ns for both cases, leaving room for further optimization of the cycle time by about 40~ns in future experiments.

We measure the basic device parameters presented in Table~\ref{table:paramp} by performing single- and two-tone spectroscopy, as well as Rabi, Ramsey and coherence measurements for the ground state to excited state and excited to second excited state transitions, as presented in Ref.~\cite{Besse2020a}. 

\section{Readout characterization and population extraction} \label{RO}
\label{sec:readout_characterization}
\begin{figure}[!t]
\includegraphics[width=\columnwidth]{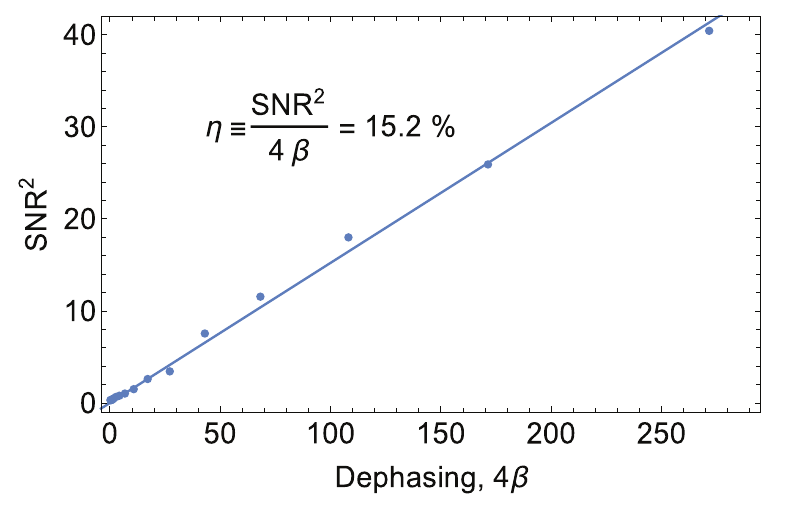}
\caption{\label{dephasing}
Squared signal-to-noise ratio $\text{SNR}^2$ vs four times the measurement induced dephasing $4 \beta$ for different readout powers. Line is a linear fit to the data, $\eta$ is extracted from the fit.}
\end{figure}

\begin{figure*}[!t]
\includegraphics[width=\textwidth]{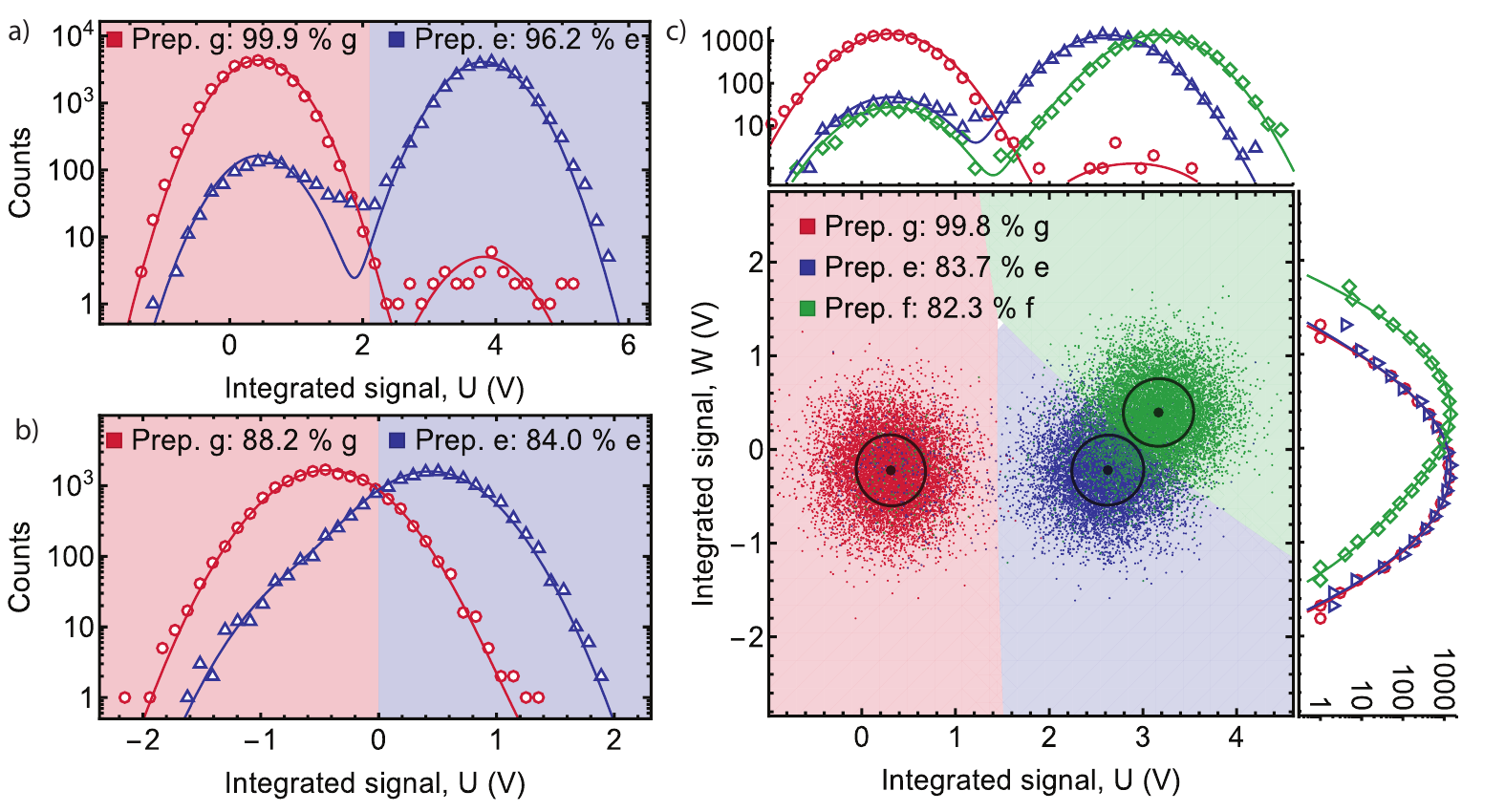}
\caption{\label{readout}
(a,b) Histogram of the integrated readout signal $U$, when preparing a ground (g, red) and excited state (e, dark blue) for (a) strong and (b) weak measurements, after heralding the ground state with a pre-selection readout pulse. Lines are a bimodal Gaussian fit to the data. (c) Two-dimensional histogram of the integrated readout signals $U$ and $W$ (see main text), when preparing a ground (g, red), excited (e, dark blue) or second excited (f, green) state after heralding the ground state with a pre-selection readout pulse. Black points and circles indicate the fitted means and standard deviation ellipses. Marginal distributions with the corresponding fits are shown in the top and right subpanel. For each panel, all prepared states are fitted with the same means and variances, but different amplitudes. Assignment regions are shown as background colors.}
\end{figure*}

To characterize the readout detection chain, we measure the dephasing $\beta = |\rho_{01,\text{on}}(T)|/|\rho_{01,\text{off}}(T)|$ induced by a readout pulse of length $T$, where $\rho_{01,\text{on}}(t)$ is time dependent off-diagonal element of the qubit's density matrix with the readout pulse present, while $\rho_{01,\text{off}}(t)$ is for without the readout pulse. We then compare $\beta$ to the signal-to-noise ratio ($\text{SNR}$) of the processed readout signal \cite{Bultink2018}, see Fig.~\ref{readout}(a). As, for a given dephasing $\beta$, $\sqrt{4\beta}$ is the maximum possible $\text{SNR}$, we define the quantum efficiency $\eta$ as $\eta =\text{SNR}^2/4 \beta$. We measure $\beta$ in a Ramsey-like experiment, applying a readout pulse of varying amplitude in between the two $\pi/2$ pulses. For these amplitudes, we then also evaluate the $\text{SNR}$ of the measurement signal, by preparing the qubit in $\ket{g}$ and $\ket{e}$, creating a histogram of the integrated signal $U$ and fitting a bimodal Gaussian distribution $a_g \mathcal{N}(\mu_g,\sigma_g^2) + a_e \mathcal{N}(\mu_e,\sigma_e^2)$, with means $\mu_g$, $\mu_e$, variances $\sigma_g^2,\sigma_e^2$ and amplitudes $a_g$,$a_e$ to $U$. The SNR is then defined as $\text{SNR}^2=|\mu_g-\mu_e|^2/\sigma_g^2$. We observe, as expected, a linear dependence between $\text{SNR}^2$ and $4 \beta$, see Fig.~\ref{dephasing}, and we obtain the quantum efficiency of $\eta=15.2\%$ from a linear fit to the data, likely due to losses before the TWPA and added noise by amplifiers after the TWPA, whose gain was not large enough to overcome other noise sources.

Furthermore, we evaluate the performance of the readout in different regimes by extracting the readout infidelity $1-\mathcal{F}$. For this purpose, we prepare the qubit in $\ket{g}$, $\ket{e}$ (and second excited $\ket{f}$) states, after heralding the ground state with a pre-selection readout pulse, and fit a bimodal (trimodal) Gaussian distribution to the combined histogram of both (all) prepared states, see Fig.~\ref{readout}.
For two-level-readout, we define a threshold $t=(\mu_g + \mu_e)/2$ and assign shots with $U<t$ to $\ket{g}$ and with $U>t$ to $\ket{e}$, see Fig.~\ref{readout}(a,b). By counting the missassigned shots, we extract $P(g|e)$, the probability to assign a prepared excited state to $\ket{g}$, and $P(e|g)$ the probability to assign a prepared ground state to $\ket{e}$, and obtain the readout infidelity of $1-\mathcal{F}=\frac{1}{2} \left(P(g|e) + P(e|g) \right)=1.95$~$\%$ for strong and $13.9$~$\%$ for weak measurements. The strong measurements are limited by the decay of the excited state into the ground state during the 256~ns-long readout pulse, while overlap errors dominate in the weak measurement case. For three-level readout, we define three assignment regions based on the fitted Gaussian distributions and obtain an infidelity of $1-\mathcal{F}=11.3$~$\%$, see Fig.~\ref{readout}(c). We note that the readout was optimized for two-level readout, resulting in the comparatively large error \cite{Magnard2018,Krinner2022} when choosing to distinguish between all three states.

These finite readout infidelities $1-\mathcal{F}$ will lead to errors in the extraction of the initialization error $1-P_g$ when using thresholding. Therefore we use a a different method: We first obtain the means $\mu_g$ and $\mu_e$ and variances $\sigma_g^2$ and $\sigma_e^2$ from a fit of a bimodal Gaussian distribution to the histogram of the initial equilibrium state. For the weak measurement case, we facilitate the fitting by assuming $\sigma_g^2=\sigma_e^2$, as this is expected for low readout powers. In the three-level case, we fit a two-dimensional tri-modal Gaussian distribution $a_g \mathcal{N}(\mu_g,\Sigma_g) + a_e \mathcal{N}(\mu_e,\Sigma_e) + a_f \mathcal{N}(\mu_f,\Sigma_f)$ to the two-dimensional histogram of $U$ and $W$ of the initial equilibrium state, and obtain means $\mu_g$, $\mu_e$ and $\mu_f$ and covariance matrices $\Sigma_g$, $\Sigma_e$ and $\Sigma_f$. In a second step, we then fit the amplitudes $a_g$ and $a_e$ ($a_f$) to the histogram of $U$ (and $W$) from the verification measurement, using the previously obtained means and variances (covariance matrices). The extracted populations are then given by the amplitude ratios $P_g=a_g/(a_g+a_e)$ and $P_e=a_e/(a_g+a_e)$ ($P_g=a_g/(a_g+a_e+a_f)$, $P_e=a_e/(a_g+a_e+a_g)$ and $P_f=a_e/(a_g+a_e+a_f)$). We note that the binning of the shots in the histogram leads to Poissonian noise, making standard least squares fitting procedures inaccurate. Instead we use a maximum likelihood procedure as described in Ref.~\cite{Laurence2010} to fit the bi-/tri-modal Gaussian distributions.

\section{State discrimination with neural networks} \label{Disc}
Since qubit state initialization relies on the distinguishability between the two states given an observation $\vectorbold{s}$, we also study the ability of the neural network to accomplish this task. We compare its performance with the one of a standard classifier, which integrates $\vectorbold{s}$ with a set of optimal filter coefficients $\vectorbold{w_s}$ to obtain $U= \vectorbold{s} \cdot \vectorbold{w_s}$ and assigns a state by thresholding $U$ \cite{Gambetta2007,Walter2017}. To train the neural network in assigning the correct state, we use supervised learning \cite{Lienhard2022} on a labelled data set consisting of 8212 individual time-traces in which we prepare ground and excited states after heralding an initial ground state with a pre-selection readout. The performance of the neural network classifier is evaluated based on an independent validation data set, which was interleaved with the training data set.

\begin{figure}[!t]
\includegraphics[width=\columnwidth]{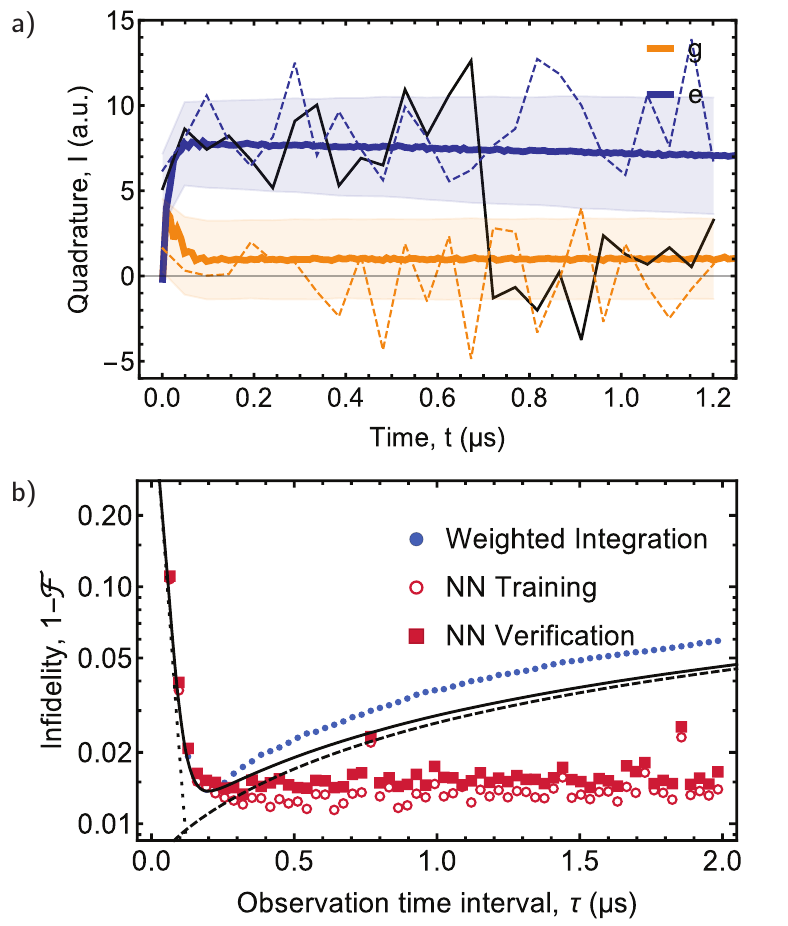}
\caption{\label{SL}
State discrimination via neural network. (a) Measured average (solid lines, $\pm \sigma$ standard deviation shaded) and single-shot (dotted) quadrature $I$ (single-shot down-sampled with a six point boxcar filter) for the qubit being in the ground (orange) or the excited (dark blue) state, as well as a single-shot during a potential decay event (black). (b) Readout infidelity $1-\mathcal{F}$ with respect to the prepared state vs observation time $\tau$, when assigning the states with the standard classifier (blue dots) and with a trained neural network classifier for a verification data set (red squares) and a training data set (red circles), respectively. Simulated readout infidelity of the standard classifier (solid black line), considering overlap errors (dotted) and errors due to decoherence (dashed).
}
\end{figure}

The two example time-traces for prepared $\ket{g}$ and $\ket{e}$ states (dashed blue and orange lines in Fig.~\ref{SL}(a)) become distinguishable on a timescale of about 50~ns. The fluctuations around their respective average response $\langle \vectorbold{s}_g \rangle$ and $\langle \vectorbold{s}_e \rangle$ (solid orange and blue lines) are dominated by Gaussian noise added during the amplification process. In addition, there are few instances in which the time-dependent signal suddenly changes its amplitude (black trace in Fig.~\ref{SL}(a)), indicating possible state transition events during the measurement mostly due to decay from $\ket{e}$ to $\ket{g}$.

For short measurement times $\tau$ up to $200$~ns the readout infidelity $1-\mathcal{F}=\frac{1}{2} \left(P(g|e) + P(e|g) \right)$, where $P(i|j)$ is the fraction of states prepared in state $\ket{j}$ and assigned to $\ket{i}$, decreases for both classifiers when increasing $\tau$, see Fig.~\ref{SL}(b). The neural network's performance matches the readout fidelity of the standard classifier, which is known to be optimal for integration times much shorter than the qubit lifetime $\tau \ll T_1$. For longer measurement times $\tau$, the readout infidelity of the standard classifier starts to increase because of state transitions during the measurement, which the linear filtering technique cannot resolve. In contrast, the neural network's performance further improves up until $\tau \approx 300$~ns and stays constant afterwards as the neural network is able to detect such state transition events. Thus the neural network outperforms the standard classifier in this regime. For example, the neural network classifier correctly assigned the time-trace shown in black in Fig.~\ref{SL}(a), while it was misclassified by the standard classifier. We note that  for the two outliers in the performance of the neural network (around $\tau=0.75$~$\SIUnitSymbolMicro$s and $\tau=1.8$~$\SIUnitSymbolMicro$s in Fig.~\ref{SL}(b)) the training algorithm most likely converged to non-optimal network parameters.

\section{Experimental reinforcement learning}
\label{Train}

\begin{algorithm}
	\begin{algorithmic}
		\Require Initial network parameters $\theta$ and $\zeta$ of the policy and critic network $\pi_\theta$ and $V_\zeta$
		\For{training step=1,2,...,$N_\mathrm{steps}$}
		\State Transfer $\theta$ to the FPGA
		\State Record episodes in the quantum system
		\State Transfer episodes to the PC
		\State Evaluate the critic network $V_\zeta$
		\State Calculate the rewards (see \cref{equ:used_reward})
		\State Update $\theta$ and $\zeta$ with PPO \cite{schulman_proximal_2017, hill_stable_2018}
		\EndFor
	\end{algorithmic}
	\caption{Training algorithm}\label{alg:training}
\end{algorithm}

\begin{table}
	\centering
	\begin{tabular}{l c} 
		\hline
		\textbf{Hyperparameter} & \textbf{Value} \\
		\hline
		Adam parameter $\eta$ & \num{5e-4}  \\ 
		Adam parameter $\beta_1$ & \num{0.98} \\
		Adam parameter $\beta_2$ & \num{0.999} \\
		Discount rate $\gamma$ & \num{0.92} \\
		Entropy coefficient & \num{0.01} \\
		Cliprange & \num{0.04} \\
		$\lambda$ of the generalized advantage estimation & \num{0.98} \\
		Number of training minibatches per update & \num{1} \\
		Number of epochs for surrogate optimization & \num{8}  \\
		Maximum value for gradient clipping & $\infty$  \\
		\hline
	\end{tabular}
	\caption{\label{tab:hyperparameters} Hyperparameters used for training, for definition of the hyperparameters see \cite{hill_stable_2018, schulman_proximal_2017}}
\end{table}

To train the network, we perform 500 training steps, i.e. 500 updates of the neural network parameters $\theta$, in around \SI{8}{\minute}, starting from a random policy. During each training step, the FPGA records initialization episodes with a cycle time of \SI{856}{\nano\second} and a repetition rate of $10$~KHz until 1000 measurements have been carried out. The measurement outcomes and the chosen actions are transferred to a personal computer (PC) in around \SI{0.6}{\second}. The PC updates the parameters $\theta$ of the policy network $\pi_\theta$ within \SI{0.1}{\second}, and the updated parameters $\theta$ are transferred back to the FPGA in around \SI{0.3}{\second}.

The only accessible information during the initialization procedure is the measurement results during the feedback loop and the verification measurements; thus, the reward function can only be based on them. We choose the  reward $r^t$ within time step $t\in\{1,...,n\}$ as
\begin{equation}
	\label{equ:used_reward}
	r^t = \frac{U_{t+1}-U_t}{U_g-U_e} - \lambda
\end{equation}
with the projected measurement result of the $t$\textsuperscript{th} iteration $U_t$, the projected verification measurement $U_{n+1}$ and a control parameter $\lambda$. $U_g$ and $U_e$ are the average projected readout signals if the qubit is prepared in the ground or excited state. $U_{t+1}-U_t$ provides the progress of the initialization compared to the previous round and gives direct information if the action $a_t$ resulted in a quantum state closer to the target state. The parameter $\lambda$ penalizes every action and thus controls the trade-off between average episode length and initialization fidelity.

The network parameters $\theta$ are modified in every update step to maximize the averaged cumulative reward $\langle R \rangle$, defined as
\begin{equation}
	\label{equ:return_def}
	\langle R \rangle = \Big\langle \sum_{t=1}^{n} r^t \Big\rangle
\end{equation}
where $\langle \cdot \rangle$ denotes the average over all possible episodes. With our chosen reward function, the average cumulative reward equals
\begin{equation}
	\label{equ:target_return}
	\langle R \rangle = \frac{\langle U_\mathrm{ver} \rangle-U_e}{U_g - U_e} -\lambda \langle n\rangle + \mathrm{const.}
\end{equation}
The first term approximates the initialization fidelity, the second one penalizes long episodes and the constant is independent of the agent's policy.

For the training step we use the Proximal Policy Optimization (PPO) algorithm \cite{schulman_proximal_2017} from the Python library Stable Baselines \cite{hill_stable_2018}. In addition to the policy network on the FPGA, the PPO algorithm makes use of a second network, the so-called critic network $V_\zeta$ with its parameters $\zeta$. Based on the current observation, the critic estimates the expected future cumulative reward given the current policy $\pi_\theta$. By comparing the cumulative reward for each observation collected on the FPGA to the expectation of the critic, the PPO algorithm identifies action sequences that perform better than expected and modifies the agent's policy such that the agent is more likely to select these action sequences. The critic is only required for the update step while it is not required for the decision-making process. Therefore, the critic network is only running on the PC. We implement the critic network as a feedforward network with two hidden layers with 64 neurons per layer. All additional hyperparameters of the PPO algorithm are listed in \cref{tab:hyperparameters}.

The whole training loop described above is summarized in \cref{alg:training}.

\section{Low-latency neural network implemented on the FPGA}
\label{sec:min_latency_network}

In implementing the agent's policy as a neural network on an FPGA (see Fig.~3), we aimed for an architecture which achieves high initialization fidelities while keeping processing latencies at a minimum. In the following, we discuss design considerations of the neural network architecture to reach this goal by making optimal use of the available FPGA resources.

\subsection{Implementation of dense layers}
Our network is a feedforward network and consists of multiple dense layers, each of which transforms the values of $N$ input neurons $\vectorbold{y}^{(\mathrm{in})}$ into the values of $M$ output neurons $\vectorbold{y}^{(\mathrm{out})}$ according to
\begin{equation}
	y^{(\mathrm{out})}_j = f \left( \sum_{k=0}^{N-1} w_{jk} y^{(\mathrm{in})}_k + b_j \right)
\end{equation}
with $j \in \{0,...,M-1\}$. Here, $f$ is the nonlinear activation function, $w$ the $M\cross N$-dimensional kernel matrix and $\vectorbold{b}$ the $M$-dimensional bias vector (where $w$ and $\vectorbold{b}$ are different for different layers).

\begin{figure}[H]
	\includegraphics[width=\columnwidth]{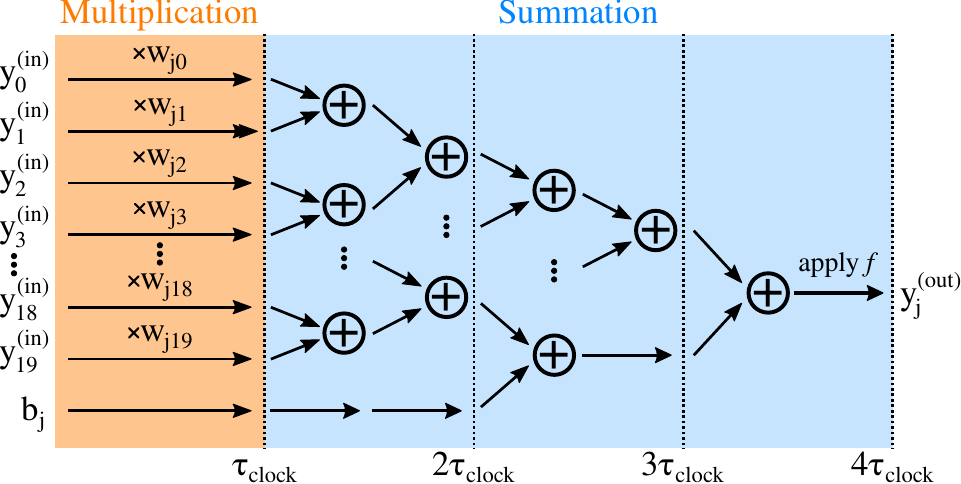}
	\caption{\label{fig:evaluation_dense_layer} Evaluation for $N=20$ input neurons on the FPGA. All output neurons are evaluated in parallel. In the first clock cycle, the inputs are multiplied with their respective weights. In the following cycles, these products and the bias are summed pairwise. Two summations are performed per clock cycle.}
\end{figure}

To compute the values of all $y^{(\mathrm{out})}_j$ in parallel and with minimum latency on the FPGA we multiply all $y^{(\mathrm{in})}_k$ with their respective weights $w_{jk}$ within one clock cycle of duration $\tau_\mathrm{clock}=\SI{8}{\nano\second}$ and then add them up sequentially  in subsequent clock cycles, see \cref{fig:evaluation_dense_layer}.  Since two subsequent additions are performed within one clock cycle,  the number of summands gets reduced in each clock cycle by at most a factor $2^2=4$, such that the total number of clock cycles required to perform the summation is $\lceil\log_4(N+1)\rceil$, where the  "+1" accounts for the bias $b_j$. We evaluate the nonlinear activation function $f$ chosen to be the Rectified Linear Unit (ReLU) function in the last clock cycle. Therefore, the execution time $\tau_\mathrm{dense}$ of a single dense layer is given by
\begin{equation}
	\label{equ:execution_time_dense_layer}
	\tau_\mathrm{dense} = \tau_\mathrm{clock} \left(1+ \lceil\log_4 (N+1)\rceil \right),
\end{equation}
In our specific experiment, we choose $N=20$ for each layer of the low-latency network resulting in an execution time of $\SI{32}{\nano \second}$ per layer.

\subsection{Implementation of the preprocessing network}
We equip our network with a memory of the past by providing it in each cycle $t$ with the readout signals $s^{j}$ and actions $a^{j}$ from $l$ previous rounds.
As this information is already available after the previous action was selected, this input is evaluated in a \emph{pre-processing network} (see Fig.~3) while the agent is waiting to receive the most recent readout signal $\vectorbold{s}^{t}$ of the current cycle. Thus, no additional latency is introduced to the feedback loop. To reduce the amount of data to be processed, we apply a 32-point boxcar filter to the previous measurement results before feeding them into the network. Each of the previous actions is expressed by a three-bit string. A fourth bit is added if the \emph{gf}-flip action is considered. The preprocessing network consists of two layers with $12$ neurons per layer.

\newpage

\subsection{Implementation of the low-latency network}
We start to evaluate the network as soon as the first element of the signal arrives. The first layer processes this information together with the output of the preprocessing network. Layer by layer, the most recent measurement data is fed into the network until the whole signal is processed.

The measurement signal, recorded with a time resolution of \SI{1}{\nano\second}, is down-sampled with an eight-point boxcar filter, introducing a latency of \SI{16}{\nano\second}. We feed four elements of the in-phase and out-of-phase component of the down-sampled signal and the output of $12$ neurons from the previous layer into the subsequent layer resulting in an input size of $N=20$.

The last layer has one output neuron per action and each neuron value encodes the probability of choosing its corresponding action. In order to sample an action, we use the Gumbel-max trick \cite{huijben_review_2021} which does not introduce any additional latencies.

The network execution adds a latency of \SI{48}{\nano\second}, where  \SI{16}{\nano\second} result from the eight-point boxcar filter and \SI{32}{\nano \second} from the execution of the last layer.

%% file: RLonFPGA.bbl
\begin{thebibliography}{57}%
\makeatletter
\providecommand \@ifxundefined [1]{%
 \@ifx{#1\undefined}
}%
\providecommand \@ifnum [1]{%
 \ifnum #1\expandafter \@firstoftwo
 \else \expandafter \@secondoftwo
 \fi
}%
\providecommand \@ifx [1]{%
 \ifx #1\expandafter \@firstoftwo
 \else \expandafter \@secondoftwo
 \fi
}%
\providecommand \natexlab [1]{#1}%
\providecommand \enquote  [1]{``#1''}%
\providecommand \bibnamefont  [1]{#1}%
\providecommand \bibfnamefont [1]{#1}%
\providecommand \citenamefont [1]{#1}%
\providecommand \href@noop [0]{\@secondoftwo}%
\providecommand \href [0]{\begingroup \@sanitize@url \@href}%
\providecommand \@href[1]{\@@startlink{#1}\@@href}%
\providecommand \@@href[1]{\endgroup#1\@@endlink}%
\providecommand \@sanitize@url [0]{\catcode `\\12\catcode `\$12\catcode
  `\&12\catcode `\#12\catcode `\^12\catcode `\_12\catcode `\%12\relax}%
\providecommand \@@startlink[1]{}%
\providecommand \@@endlink[0]{}%
\providecommand \url  [0]{\begingroup\@sanitize@url \@url }%
\providecommand \@url [1]{\endgroup\@href {#1}{\urlprefix }}%
\providecommand \urlprefix  [0]{URL }%
\providecommand \Eprint [0]{\href }%
\providecommand \doibase [0]{https://doi.org/}%
\providecommand \selectlanguage [0]{\@gobble}%
\providecommand \bibinfo  [0]{\@secondoftwo}%
\providecommand \bibfield  [0]{\@secondoftwo}%
\providecommand \translation [1]{[#1]}%
\providecommand \BibitemOpen [0]{}%
\providecommand \bibitemStop [0]{}%
\providecommand \bibitemNoStop [0]{.\EOS\space}%
\providecommand \EOS [0]{\spacefactor3000\relax}%
\providecommand \BibitemShut  [1]{\csname bibitem#1\endcsname}%
\let\auto@bib@innerbib\@empty
\bibitem [{\citenamefont {Rist\`e}\ \emph {et~al.}(2012)\citenamefont
  {Rist\`e}, \citenamefont {Bultink}, \citenamefont {Lehnert},\ and\
  \citenamefont {DiCarlo}}]{Riste2012b}%
  \BibitemOpen
  \bibfield  {author} {\bibinfo {author} {\bibfnamefont {D.}~\bibnamefont
  {Rist\`e}}, \bibinfo {author} {\bibfnamefont {C.~C.}\ \bibnamefont
  {Bultink}}, \bibinfo {author} {\bibfnamefont {K.~W.}\ \bibnamefont
  {Lehnert}},\ and\ \bibinfo {author} {\bibfnamefont {L.}~\bibnamefont
  {DiCarlo}},\ }\bibfield  {title} {\bibinfo {title} {Feedback control of a
  solid-state qubit using high-fidelity projective measurement},\ }\href
  {https://doi.org/10.1103/PhysRevLett.109.240502} {\bibfield  {journal}
  {\bibinfo  {journal} {Phys. Rev. Lett.}\ }\textbf {\bibinfo {volume} {109}},\
  \bibinfo {pages} {240502} (\bibinfo {year} {2012})}\BibitemShut {NoStop}%
\bibitem [{\citenamefont {Salath\'e}\ \emph {et~al.}(2018)\citenamefont
  {Salath\'e}, \citenamefont {Kurpiers}, \citenamefont {Karg}, \citenamefont
  {Lang}, \citenamefont {Andersen}, \citenamefont {Akin}, \citenamefont
  {Krinner}, \citenamefont {Eichler},\ and\ \citenamefont
  {Wallraff}}]{Salathe2018}%
  \BibitemOpen
  \bibfield  {author} {\bibinfo {author} {\bibfnamefont {Y.}~\bibnamefont
  {Salath\'e}}, \bibinfo {author} {\bibfnamefont {P.}~\bibnamefont {Kurpiers}},
  \bibinfo {author} {\bibfnamefont {T.}~\bibnamefont {Karg}}, \bibinfo {author}
  {\bibfnamefont {C.}~\bibnamefont {Lang}}, \bibinfo {author} {\bibfnamefont
  {C.~K.}\ \bibnamefont {Andersen}}, \bibinfo {author} {\bibfnamefont
  {A.}~\bibnamefont {Akin}}, \bibinfo {author} {\bibfnamefont {S.}~\bibnamefont
  {Krinner}}, \bibinfo {author} {\bibfnamefont {C.}~\bibnamefont {Eichler}},\
  and\ \bibinfo {author} {\bibfnamefont {A.}~\bibnamefont {Wallraff}},\
  }\bibfield  {title} {\bibinfo {title} {Low-latency digital signal processing
  for feedback and feedforward in quantum computing and communication},\ }\href
  {https://doi.org/10.1103/PhysRevApplied.9.034011} {\bibfield  {journal}
  {\bibinfo  {journal} {Phys. Rev. Appl.}\ }\textbf {\bibinfo {volume} {9}},\
  \bibinfo {pages} {034011} (\bibinfo {year} {2018})}\BibitemShut {NoStop}%
\bibitem [{\citenamefont {Negnevitsky}\ \emph {et~al.}(2018)\citenamefont
  {Negnevitsky}, \citenamefont {Marinelli}, \citenamefont {Mehta},
  \citenamefont {Lo}, \citenamefont {Flühmann},\ and\ \citenamefont
  {Home}}]{Negnevitsky2018}%
  \BibitemOpen
  \bibfield  {author} {\bibinfo {author} {\bibfnamefont {V.}~\bibnamefont
  {Negnevitsky}}, \bibinfo {author} {\bibfnamefont {M.}~\bibnamefont
  {Marinelli}}, \bibinfo {author} {\bibfnamefont {K.~K.}\ \bibnamefont
  {Mehta}}, \bibinfo {author} {\bibfnamefont {H.-Y.}\ \bibnamefont {Lo}},
  \bibinfo {author} {\bibfnamefont {C.}~\bibnamefont {Flühmann}},\ and\
  \bibinfo {author} {\bibfnamefont {J.~P.}\ \bibnamefont {Home}},\ }\bibfield
  {title} {\bibinfo {title} {Repeated multi-qubit readout and feedback with a
  mixed-species trapped-ion register},\ }\href
  {https://doi.org/10.1038/s41586-018-0668-z} {\bibfield  {journal} {\bibinfo
  {journal} {Nature}\ }\textbf {\bibinfo {volume} {563}},\ \bibinfo {pages}
  {527} (\bibinfo {year} {2018})}\BibitemShut {NoStop}%
\bibitem [{\citenamefont {Steffen}\ \emph {et~al.}(2013)\citenamefont
  {Steffen}, \citenamefont {Salathe}, \citenamefont {Oppliger}, \citenamefont
  {Kurpiers}, \citenamefont {Baur}, \citenamefont {Lang}, \citenamefont
  {Eichler}, \citenamefont {Puebla-Hellmann}, \citenamefont {Fedorov},\ and\
  \citenamefont {Wallraff}}]{Steffen2013}%
  \BibitemOpen
  \bibfield  {author} {\bibinfo {author} {\bibfnamefont {L.}~\bibnamefont
  {Steffen}}, \bibinfo {author} {\bibfnamefont {Y.}~\bibnamefont {Salathe}},
  \bibinfo {author} {\bibfnamefont {M.}~\bibnamefont {Oppliger}}, \bibinfo
  {author} {\bibfnamefont {P.}~\bibnamefont {Kurpiers}}, \bibinfo {author}
  {\bibfnamefont {M.}~\bibnamefont {Baur}}, \bibinfo {author} {\bibfnamefont
  {C.}~\bibnamefont {Lang}}, \bibinfo {author} {\bibfnamefont {C.}~\bibnamefont
  {Eichler}}, \bibinfo {author} {\bibfnamefont {G.}~\bibnamefont
  {Puebla-Hellmann}}, \bibinfo {author} {\bibfnamefont {A.}~\bibnamefont
  {Fedorov}},\ and\ \bibinfo {author} {\bibfnamefont {A.}~\bibnamefont
  {Wallraff}},\ }\bibfield  {title} {\bibinfo {title} {Deterministic quantum
  teleportation with feed-forward in a solid state system},\ }\href
  {https://doi.org/10.1038/nature12422} {\bibfield  {journal} {\bibinfo
  {journal} {Nature}\ }\textbf {\bibinfo {volume} {500}},\ \bibinfo {pages}
  {319} (\bibinfo {year} {2013})}\BibitemShut {NoStop}%
\bibitem [{\citenamefont {{Chou}}\ \emph {et~al.}(2018)\citenamefont {{Chou}},
  \citenamefont {{Blumoff}}, \citenamefont {{Wang}}, \citenamefont
  {{Reinhold}}, \citenamefont {{Axline}}, \citenamefont {{Gao}}, \citenamefont
  {{Frunzio}}, \citenamefont {{Devoret}}, \citenamefont {{Jiang}},\ and\
  \citenamefont {{Schoelkopf}}}]{Chou2018}%
  \BibitemOpen
  \bibfield  {author} {\bibinfo {author} {\bibfnamefont {K.~S.}\ \bibnamefont
  {{Chou}}}, \bibinfo {author} {\bibfnamefont {J.~Z.}\ \bibnamefont
  {{Blumoff}}}, \bibinfo {author} {\bibfnamefont {C.~S.}\ \bibnamefont
  {{Wang}}}, \bibinfo {author} {\bibfnamefont {P.~C.}\ \bibnamefont
  {{Reinhold}}}, \bibinfo {author} {\bibfnamefont {C.~J.}\ \bibnamefont
  {{Axline}}}, \bibinfo {author} {\bibfnamefont {Y.~Y.}\ \bibnamefont {{Gao}}},
  \bibinfo {author} {\bibfnamefont {L.}~\bibnamefont {{Frunzio}}}, \bibinfo
  {author} {\bibfnamefont {M.~H.}\ \bibnamefont {{Devoret}}}, \bibinfo {author}
  {\bibfnamefont {L.}~\bibnamefont {{Jiang}}},\ and\ \bibinfo {author}
  {\bibfnamefont {R.~J.}\ \bibnamefont {{Schoelkopf}}},\ }\bibfield  {title}
  {\bibinfo {title} {{Deterministic teleportation of a quantum gate between two
  logical qubits}},\ }\href {https://doi.org/10.1038/s41586-018-0470-y}
  {\bibfield  {journal} {\bibinfo  {journal} {Nature}\ }\textbf {\bibinfo
  {volume} {561}},\ \bibinfo {pages} {368} (\bibinfo {year}
  {2018})}\BibitemShut {NoStop}%
\bibitem [{\citenamefont {Ofek}\ \emph {et~al.}(2016)\citenamefont {Ofek},
  \citenamefont {Petrenko}, \citenamefont {Heeres}, \citenamefont {Reinhold},
  \citenamefont {Leghtas}, \citenamefont {Vlastakis}, \citenamefont {Liu},
  \citenamefont {Frunzio}, \citenamefont {Girvin}, \citenamefont {Jiang},
  \citenamefont {Mirrahimi}, \citenamefont {Devoret},\ and\ \citenamefont
  {Schoelkopf}}]{Ofek2016}%
  \BibitemOpen
  \bibfield  {author} {\bibinfo {author} {\bibfnamefont {N.}~\bibnamefont
  {Ofek}}, \bibinfo {author} {\bibfnamefont {A.}~\bibnamefont {Petrenko}},
  \bibinfo {author} {\bibfnamefont {R.}~\bibnamefont {Heeres}}, \bibinfo
  {author} {\bibfnamefont {P.}~\bibnamefont {Reinhold}}, \bibinfo {author}
  {\bibfnamefont {Z.}~\bibnamefont {Leghtas}}, \bibinfo {author} {\bibfnamefont
  {B.}~\bibnamefont {Vlastakis}}, \bibinfo {author} {\bibfnamefont
  {Y.}~\bibnamefont {Liu}}, \bibinfo {author} {\bibfnamefont {L.}~\bibnamefont
  {Frunzio}}, \bibinfo {author} {\bibfnamefont {S.~M.}\ \bibnamefont {Girvin}},
  \bibinfo {author} {\bibfnamefont {L.}~\bibnamefont {Jiang}}, \bibinfo
  {author} {\bibfnamefont {M.}~\bibnamefont {Mirrahimi}}, \bibinfo {author}
  {\bibfnamefont {M.~H.}\ \bibnamefont {Devoret}},\ and\ \bibinfo {author}
  {\bibfnamefont {R.~J.}\ \bibnamefont {Schoelkopf}},\ }\bibfield  {title}
  {\bibinfo {title} {Extending the lifetime of a quantum bit with error
  correction in superconducting circuits},\ }\href
  {http://dx.doi.org/10.1038/nature18949} {\bibfield  {journal} {\bibinfo
  {journal} {Nature}\ }\textbf {\bibinfo {volume} {536}},\ \bibinfo {pages}
  {441} (\bibinfo {year} {2016})}\BibitemShut {NoStop}%
\bibitem [{\citenamefont {Andersen}\ \emph {et~al.}(2019)\citenamefont
  {Andersen}, \citenamefont {Remm}, \citenamefont {Lazar}, \citenamefont
  {Krinner}, \citenamefont {Heinsoo}, \citenamefont {Besse}, \citenamefont
  {Gabureac}, \citenamefont {Wallraff},\ and\ \citenamefont
  {Eichler}}]{Andersen2019}%
  \BibitemOpen
  \bibfield  {author} {\bibinfo {author} {\bibfnamefont {C.~K.}\ \bibnamefont
  {Andersen}}, \bibinfo {author} {\bibfnamefont {A.}~\bibnamefont {Remm}},
  \bibinfo {author} {\bibfnamefont {S.}~\bibnamefont {Lazar}}, \bibinfo
  {author} {\bibfnamefont {S.}~\bibnamefont {Krinner}}, \bibinfo {author}
  {\bibfnamefont {J.}~\bibnamefont {Heinsoo}}, \bibinfo {author} {\bibfnamefont
  {J.-C.}\ \bibnamefont {Besse}}, \bibinfo {author} {\bibfnamefont
  {M.}~\bibnamefont {Gabureac}}, \bibinfo {author} {\bibfnamefont
  {A.}~\bibnamefont {Wallraff}},\ and\ \bibinfo {author} {\bibfnamefont
  {C.}~\bibnamefont {Eichler}},\ }\bibfield  {title} {\bibinfo {title}
  {Entanglement stabilization using ancilla-based parity detection and
  real-time feedback in superconducting circuits},\ }\href
  {https://doi.org/10.1038/s41534-019-0185-4} {\bibfield  {journal} {\bibinfo
  {journal} {npj Quantum Information}\ }\textbf {\bibinfo {volume} {5}},\
  \bibinfo {pages} {69} (\bibinfo {year} {2019})}\BibitemShut {NoStop}%
\bibitem [{\citenamefont {Ryan-Anderson}\ \emph {et~al.}(2021)\citenamefont
  {Ryan-Anderson}, \citenamefont {Bohnet}, \citenamefont {Lee}, \citenamefont
  {Gresh}, \citenamefont {Hankin}, \citenamefont {Gaebler}, \citenamefont
  {Francois}, \citenamefont {Chernoguzov}, \citenamefont {Lucchetti},
  \citenamefont {Brown}, \citenamefont {Gatterman}, \citenamefont {Halit},
  \citenamefont {Gilmore}, \citenamefont {Gerber}, \citenamefont {Neyenhuis},
  \citenamefont {Hayes},\ and\ \citenamefont {Stutz}}]{RyanAnderson2021}%
  \BibitemOpen
  \bibfield  {author} {\bibinfo {author} {\bibfnamefont {C.}~\bibnamefont
  {Ryan-Anderson}}, \bibinfo {author} {\bibfnamefont {J.~G.}\ \bibnamefont
  {Bohnet}}, \bibinfo {author} {\bibfnamefont {K.}~\bibnamefont {Lee}},
  \bibinfo {author} {\bibfnamefont {D.}~\bibnamefont {Gresh}}, \bibinfo
  {author} {\bibfnamefont {A.}~\bibnamefont {Hankin}}, \bibinfo {author}
  {\bibfnamefont {J.~P.}\ \bibnamefont {Gaebler}}, \bibinfo {author}
  {\bibfnamefont {D.}~\bibnamefont {Francois}}, \bibinfo {author}
  {\bibfnamefont {A.}~\bibnamefont {Chernoguzov}}, \bibinfo {author}
  {\bibfnamefont {D.}~\bibnamefont {Lucchetti}}, \bibinfo {author}
  {\bibfnamefont {N.~C.}\ \bibnamefont {Brown}}, \bibinfo {author}
  {\bibfnamefont {T.~M.}\ \bibnamefont {Gatterman}}, \bibinfo {author}
  {\bibfnamefont {S.~K.}\ \bibnamefont {Halit}}, \bibinfo {author}
  {\bibfnamefont {K.}~\bibnamefont {Gilmore}}, \bibinfo {author} {\bibfnamefont
  {J.~A.}\ \bibnamefont {Gerber}}, \bibinfo {author} {\bibfnamefont
  {B.}~\bibnamefont {Neyenhuis}}, \bibinfo {author} {\bibfnamefont
  {D.}~\bibnamefont {Hayes}},\ and\ \bibinfo {author} {\bibfnamefont {R.~P.}\
  \bibnamefont {Stutz}},\ }\bibfield  {title} {\bibinfo {title} {Realization of
  real-time fault-tolerant quantum error correction},\ }\href
  {https://doi.org/10.1103/PhysRevX.11.041058} {\bibfield  {journal} {\bibinfo
  {journal} {Phys. Rev. X}\ }\textbf {\bibinfo {volume} {11}},\ \bibinfo
  {pages} {041058} (\bibinfo {year} {2021})}\BibitemShut {NoStop}%
\bibitem [{\citenamefont {Sutton}\ and\ \citenamefont
  {Barto}(2018)}]{sutton2018reinforcement}%
  \BibitemOpen
  \bibfield  {author} {\bibinfo {author} {\bibfnamefont {R.~S.}\ \bibnamefont
  {Sutton}}\ and\ \bibinfo {author} {\bibfnamefont {A.~G.}\ \bibnamefont
  {Barto}},\ }\href@noop {} {\emph {\bibinfo {title} {Reinforcement learning:
  An introduction}}}\ (\bibinfo  {publisher} {MIT press},\ \bibinfo {year}
  {2018})\BibitemShut {NoStop}%
\bibitem [{\citenamefont {Silver}\ \emph {et~al.}(2016)\citenamefont {Silver},
  \citenamefont {Huang}, \citenamefont {Maddison}, \citenamefont {Guez},
  \citenamefont {Sifre}, \citenamefont {Van Den~Driessche}, \citenamefont
  {Schrittwieser}, \citenamefont {Antonoglou}, \citenamefont {Panneershelvam},
  \citenamefont {Lanctot} \emph {et~al.}}]{silver2016mastering}%
  \BibitemOpen
  \bibfield  {author} {\bibinfo {author} {\bibfnamefont {D.}~\bibnamefont
  {Silver}}, \bibinfo {author} {\bibfnamefont {A.}~\bibnamefont {Huang}},
  \bibinfo {author} {\bibfnamefont {C.~J.}\ \bibnamefont {Maddison}}, \bibinfo
  {author} {\bibfnamefont {A.}~\bibnamefont {Guez}}, \bibinfo {author}
  {\bibfnamefont {L.}~\bibnamefont {Sifre}}, \bibinfo {author} {\bibfnamefont
  {G.}~\bibnamefont {Van Den~Driessche}}, \bibinfo {author} {\bibfnamefont
  {J.}~\bibnamefont {Schrittwieser}}, \bibinfo {author} {\bibfnamefont
  {I.}~\bibnamefont {Antonoglou}}, \bibinfo {author} {\bibfnamefont
  {V.}~\bibnamefont {Panneershelvam}}, \bibinfo {author} {\bibfnamefont
  {M.}~\bibnamefont {Lanctot}}, \emph {et~al.},\ }\bibfield  {title} {\bibinfo
  {title} {Mastering the game of {G}o with deep neural networks and tree
  search},\ }\href {https://doi.org/10.1038/nature16961} {\bibfield  {journal}
  {\bibinfo  {journal} {Nature}\ }\textbf {\bibinfo {volume} {529}},\ \bibinfo
  {pages} {484} (\bibinfo {year} {2016})}\BibitemShut {NoStop}%
\bibitem [{\citenamefont {Kober}\ \emph {et~al.}(2013)\citenamefont {Kober},
  \citenamefont {Bagnell},\ and\ \citenamefont
  {Peters}}]{kober2013reinforcement}%
  \BibitemOpen
  \bibfield  {author} {\bibinfo {author} {\bibfnamefont {J.}~\bibnamefont
  {Kober}}, \bibinfo {author} {\bibfnamefont {J.~A.}\ \bibnamefont {Bagnell}},\
  and\ \bibinfo {author} {\bibfnamefont {J.}~\bibnamefont {Peters}},\
  }\bibfield  {title} {\bibinfo {title} {{Reinforcement learning in robotics: A
  survey}},\ }\href {https://doi.org/10.1177/0278364913495721} {\bibfield
  {journal} {\bibinfo  {journal} {The International Journal of Robotics
  Research}\ }\textbf {\bibinfo {volume} {32}},\ \bibinfo {pages} {1238}
  (\bibinfo {year} {2013})}\BibitemShut {NoStop}%
\bibitem [{\citenamefont {Bellemare}\ \emph {et~al.}(2020)\citenamefont
  {Bellemare}, \citenamefont {Candido}, \citenamefont {Castro}, \citenamefont
  {Gong}, \citenamefont {Machado}, \citenamefont {Moitra}, \citenamefont
  {Ponda},\ and\ \citenamefont {Wang}}]{Bellemare2020}%
  \BibitemOpen
  \bibfield  {author} {\bibinfo {author} {\bibfnamefont {M.~G.}\ \bibnamefont
  {Bellemare}}, \bibinfo {author} {\bibfnamefont {S.}~\bibnamefont {Candido}},
  \bibinfo {author} {\bibfnamefont {P.~S.}\ \bibnamefont {Castro}}, \bibinfo
  {author} {\bibfnamefont {J.}~\bibnamefont {Gong}}, \bibinfo {author}
  {\bibfnamefont {M.~C.}\ \bibnamefont {Machado}}, \bibinfo {author}
  {\bibfnamefont {S.}~\bibnamefont {Moitra}}, \bibinfo {author} {\bibfnamefont
  {S.~S.}\ \bibnamefont {Ponda}},\ and\ \bibinfo {author} {\bibfnamefont
  {Z.}~\bibnamefont {Wang}},\ }\bibfield  {title} {\bibinfo {title} {Autonomous
  navigation of stratospheric balloons using reinforcement learning},\ }\href
  {https://www.nature.com/articles/s41586-020-2939-8} {\bibfield  {journal}
  {\bibinfo  {journal} {Nature}\ }\textbf {\bibinfo {volume} {588}},\ \bibinfo
  {pages} {77} (\bibinfo {year} {2020})}\BibitemShut {NoStop}%
\bibitem [{\citenamefont {Praeger}\ \emph {et~al.}(2021)\citenamefont
  {Praeger}, \citenamefont {Xie}, \citenamefont {Grant-Jacob}, \citenamefont
  {Eason},\ and\ \citenamefont {Mills}}]{praeger2021playing}%
  \BibitemOpen
  \bibfield  {author} {\bibinfo {author} {\bibfnamefont {M.}~\bibnamefont
  {Praeger}}, \bibinfo {author} {\bibfnamefont {Y.}~\bibnamefont {Xie}},
  \bibinfo {author} {\bibfnamefont {J.~A.}\ \bibnamefont {Grant-Jacob}},
  \bibinfo {author} {\bibfnamefont {R.~W.}\ \bibnamefont {Eason}},\ and\
  \bibinfo {author} {\bibfnamefont {B.}~\bibnamefont {Mills}},\ }\bibfield
  {title} {\bibinfo {title} {Playing optical tweezers with deep reinforcement
  learning: in virtual, physical and augmented environments},\ }\href
  {https://doi.org/10.1088/2632-2153/abf0f6} {\bibfield  {journal} {\bibinfo
  {journal} {Machine Learning: Science and Technology}\ }\textbf {\bibinfo
  {volume} {2}},\ \bibinfo {pages} {035024} (\bibinfo {year}
  {2021})}\BibitemShut {NoStop}%
\bibitem [{\citenamefont {Guo}\ \emph {et~al.}(2021)\citenamefont {Guo},
  \citenamefont {Chen}, \citenamefont {Liu}, \citenamefont {Xue}, \citenamefont
  {Chen}, \citenamefont {Cao}, \citenamefont {Mao}, \citenamefont {Tey},\ and\
  \citenamefont {You}}]{Guo2021}%
  \BibitemOpen
  \bibfield  {author} {\bibinfo {author} {\bibfnamefont {S.-F.}\ \bibnamefont
  {Guo}}, \bibinfo {author} {\bibfnamefont {F.}~\bibnamefont {Chen}}, \bibinfo
  {author} {\bibfnamefont {Q.}~\bibnamefont {Liu}}, \bibinfo {author}
  {\bibfnamefont {M.}~\bibnamefont {Xue}}, \bibinfo {author} {\bibfnamefont
  {J.-J.}\ \bibnamefont {Chen}}, \bibinfo {author} {\bibfnamefont {J.-H.}\
  \bibnamefont {Cao}}, \bibinfo {author} {\bibfnamefont {T.-W.}\ \bibnamefont
  {Mao}}, \bibinfo {author} {\bibfnamefont {M.~K.}\ \bibnamefont {Tey}},\ and\
  \bibinfo {author} {\bibfnamefont {L.}~\bibnamefont {You}},\ }\bibfield
  {title} {\bibinfo {title} {Faster state preparation across quantum phase
  transition assisted by reinforcement learning},\ }\href
  {https://doi.org/10.1103/PhysRevLett.126.060401} {\bibfield  {journal}
  {\bibinfo  {journal} {Phys. Rev. Lett.}\ }\textbf {\bibinfo {volume} {126}},\
  \bibinfo {pages} {060401} (\bibinfo {year} {2021})}\BibitemShut {NoStop}%
\bibitem [{\citenamefont {Ai}\ \emph {et~al.}(2022)\citenamefont {Ai},
  \citenamefont {Ding}, \citenamefont {Ban}, \citenamefont {Martín-Guerrero},
  \citenamefont {Casanova}, \citenamefont {Cui}, \citenamefont {Huang},
  \citenamefont {Chen}, \citenamefont {Li},\ and\ \citenamefont
  {Guo}}]{Ai2022}%
  \BibitemOpen
  \bibfield  {author} {\bibinfo {author} {\bibfnamefont {M.-Z.}\ \bibnamefont
  {Ai}}, \bibinfo {author} {\bibfnamefont {Y.}~\bibnamefont {Ding}}, \bibinfo
  {author} {\bibfnamefont {Y.}~\bibnamefont {Ban}}, \bibinfo {author}
  {\bibfnamefont {J.~D.}\ \bibnamefont {Martín-Guerrero}}, \bibinfo {author}
  {\bibfnamefont {J.}~\bibnamefont {Casanova}}, \bibinfo {author}
  {\bibfnamefont {J.-M.}\ \bibnamefont {Cui}}, \bibinfo {author} {\bibfnamefont
  {Y.-F.}\ \bibnamefont {Huang}}, \bibinfo {author} {\bibfnamefont
  {X.}~\bibnamefont {Chen}}, \bibinfo {author} {\bibfnamefont {C.-F.}\
  \bibnamefont {Li}},\ and\ \bibinfo {author} {\bibfnamefont {G.-C.}\
  \bibnamefont {Guo}},\ }\bibfield  {title} {\bibinfo {title} {Experimentally
  realizing efficient quantum control with reinforcement learning},\ }\href
  {https://doi.org/10.1007/s11433-021-1841-2} {\bibfield  {journal} {\bibinfo
  {journal} {Science China Physics, Mechanics \& Astronomy}\ }\textbf {\bibinfo
  {volume} {65}},\ \bibinfo {pages} {250312} (\bibinfo {year}
  {2022})}\BibitemShut {NoStop}%
\bibitem [{\citenamefont {Kuprikov}\ \emph {et~al.}(2022)\citenamefont
  {Kuprikov}, \citenamefont {Kokhanovskiy}, \citenamefont {Serebrennikov},\
  and\ \citenamefont {Turitsyn}}]{kuprikov2022deep}%
  \BibitemOpen
  \bibfield  {author} {\bibinfo {author} {\bibfnamefont {E.}~\bibnamefont
  {Kuprikov}}, \bibinfo {author} {\bibfnamefont {A.}~\bibnamefont
  {Kokhanovskiy}}, \bibinfo {author} {\bibfnamefont {K.}~\bibnamefont
  {Serebrennikov}},\ and\ \bibinfo {author} {\bibfnamefont {S.}~\bibnamefont
  {Turitsyn}},\ }\bibfield  {title} {\bibinfo {title} {Deep reinforcement
  learning for self-tuning laser source of dissipative solitons},\ }\href
  {https://doi.org/10.1038/s41598-022-11274-w} {\bibfield  {journal} {\bibinfo
  {journal} {Scientific Reports}\ }\textbf {\bibinfo {volume} {12}},\ \bibinfo
  {pages} {1} (\bibinfo {year} {2022})}\BibitemShut {NoStop}%
\bibitem [{\citenamefont {Degrave}\ \emph {et~al.}(2022)\citenamefont
  {Degrave}, \citenamefont {Felici}, \citenamefont {Buchli}, \citenamefont
  {Neunert}, \citenamefont {Tracey}, \citenamefont {Carpanese}, \citenamefont
  {Ewalds}, \citenamefont {Hafner}, \citenamefont {Abdolmaleki}, \citenamefont
  {de~Las~Casas} \emph {et~al.}}]{degrave2022magnetic}%
  \BibitemOpen
  \bibfield  {author} {\bibinfo {author} {\bibfnamefont {J.}~\bibnamefont
  {Degrave}}, \bibinfo {author} {\bibfnamefont {F.}~\bibnamefont {Felici}},
  \bibinfo {author} {\bibfnamefont {J.}~\bibnamefont {Buchli}}, \bibinfo
  {author} {\bibfnamefont {M.}~\bibnamefont {Neunert}}, \bibinfo {author}
  {\bibfnamefont {B.}~\bibnamefont {Tracey}}, \bibinfo {author} {\bibfnamefont
  {F.}~\bibnamefont {Carpanese}}, \bibinfo {author} {\bibfnamefont
  {T.}~\bibnamefont {Ewalds}}, \bibinfo {author} {\bibfnamefont
  {R.}~\bibnamefont {Hafner}}, \bibinfo {author} {\bibfnamefont
  {A.}~\bibnamefont {Abdolmaleki}}, \bibinfo {author} {\bibfnamefont
  {D.}~\bibnamefont {de~Las~Casas}}, \emph {et~al.},\ }\bibfield  {title}
  {\bibinfo {title} {Magnetic control of tokamak plasmas through deep
  reinforcement learning},\ }\href {https://doi.org/10.1038/s41586-021-04301-9}
  {\bibfield  {journal} {\bibinfo  {journal} {Nature}\ }\textbf {\bibinfo
  {volume} {602}},\ \bibinfo {pages} {414} (\bibinfo {year}
  {2022})}\BibitemShut {NoStop}%
\bibitem [{\citenamefont {Peng}\ \emph {et~al.}(2022)\citenamefont {Peng},
  \citenamefont {Huang}, \citenamefont {Yin}, \citenamefont {Joseph},
  \citenamefont {Ramanathan},\ and\ \citenamefont {Cappellaro}}]{Peng2022}%
  \BibitemOpen
  \bibfield  {author} {\bibinfo {author} {\bibfnamefont {P.}~\bibnamefont
  {Peng}}, \bibinfo {author} {\bibfnamefont {X.}~\bibnamefont {Huang}},
  \bibinfo {author} {\bibfnamefont {C.}~\bibnamefont {Yin}}, \bibinfo {author}
  {\bibfnamefont {L.}~\bibnamefont {Joseph}}, \bibinfo {author} {\bibfnamefont
  {C.}~\bibnamefont {Ramanathan}},\ and\ \bibinfo {author} {\bibfnamefont
  {P.}~\bibnamefont {Cappellaro}},\ }\bibfield  {title} {\bibinfo {title} {Deep
  reinforcement learning for quantum hamiltonian engineering},\ }\href
  {https://doi.org/10.1103/PhysRevApplied.18.024033} {\bibfield  {journal}
  {\bibinfo  {journal} {Phys. Rev. Applied}\ }\textbf {\bibinfo {volume}
  {18}},\ \bibinfo {pages} {024033} (\bibinfo {year} {2022})}\BibitemShut
  {NoStop}%
\bibitem [{\citenamefont {T{\"u}nnermann}\ and\ \citenamefont
  {Shirakawa}(2019)}]{tunnermann2019deep}%
  \BibitemOpen
  \bibfield  {author} {\bibinfo {author} {\bibfnamefont {H.}~\bibnamefont
  {T{\"u}nnermann}}\ and\ \bibinfo {author} {\bibfnamefont {A.}~\bibnamefont
  {Shirakawa}},\ }\bibfield  {title} {\bibinfo {title} {Deep reinforcement
  learning for coherent beam combining applications},\ }\href
  {https://doi.org/10.1364/OE.27.024223} {\bibfield  {journal} {\bibinfo
  {journal} {Optics Express}\ }\textbf {\bibinfo {volume} {27}},\ \bibinfo
  {pages} {24223} (\bibinfo {year} {2019})}\BibitemShut {NoStop}%
\bibitem [{\citenamefont {Kain}\ \emph {et~al.}(2020)\citenamefont {Kain},
  \citenamefont {Hirlander}, \citenamefont {Goddard}, \citenamefont {Velotti},
  \citenamefont {Della~Porta}, \citenamefont {Bruchon},\ and\ \citenamefont
  {Valentino}}]{kain2020sample}%
  \BibitemOpen
  \bibfield  {author} {\bibinfo {author} {\bibfnamefont {V.}~\bibnamefont
  {Kain}}, \bibinfo {author} {\bibfnamefont {S.}~\bibnamefont {Hirlander}},
  \bibinfo {author} {\bibfnamefont {B.}~\bibnamefont {Goddard}}, \bibinfo
  {author} {\bibfnamefont {F.~M.}\ \bibnamefont {Velotti}}, \bibinfo {author}
  {\bibfnamefont {G.~Z.}\ \bibnamefont {Della~Porta}}, \bibinfo {author}
  {\bibfnamefont {N.}~\bibnamefont {Bruchon}},\ and\ \bibinfo {author}
  {\bibfnamefont {G.}~\bibnamefont {Valentino}},\ }\bibfield  {title} {\bibinfo
  {title} {{Sample-efficient reinforcement learning for CERN accelerator
  control}},\ }\href {https://doi.org/10.1103/PhysRevAccelBeams.23.124801}
  {\bibfield  {journal} {\bibinfo  {journal} {Physical Review Accelerators and
  Beams}\ }\textbf {\bibinfo {volume} {23}},\ \bibinfo {pages} {124801}
  (\bibinfo {year} {2020})}\BibitemShut {NoStop}%
\bibitem [{\citenamefont {Hirlaender}\ and\ \citenamefont
  {Bruchon}(2020)}]{hirlaender2020model}%
  \BibitemOpen
  \bibfield  {author} {\bibinfo {author} {\bibfnamefont {S.}~\bibnamefont
  {Hirlaender}}\ and\ \bibinfo {author} {\bibfnamefont {N.}~\bibnamefont
  {Bruchon}},\ }\bibfield  {title} {\bibinfo {title} {{Model-free and Bayesian
  Ensembling Model-based Deep Reinforcement Learning for Particle Accelerator
  Control Demonstrated on the FERMI FEL}},\ }\href
  {https://doi.org/10.48550/arXiv.2012.09737} {\bibfield  {journal} {\bibinfo
  {journal} {arXiv:2012.09737}\ } (\bibinfo {year} {2020})}\BibitemShut
  {NoStop}%
\bibitem [{\citenamefont {Yan}\ \emph {et~al.}(2021)\citenamefont {Yan},
  \citenamefont {Deng}, \citenamefont {Zhang}, \citenamefont {Zhu},
  \citenamefont {Yin}, \citenamefont {Li}, \citenamefont {Wu},\ and\
  \citenamefont {Jiang}}]{yan2021low}%
  \BibitemOpen
  \bibfield  {author} {\bibinfo {author} {\bibfnamefont {Q.}~\bibnamefont
  {Yan}}, \bibinfo {author} {\bibfnamefont {Q.}~\bibnamefont {Deng}}, \bibinfo
  {author} {\bibfnamefont {J.}~\bibnamefont {Zhang}}, \bibinfo {author}
  {\bibfnamefont {Y.}~\bibnamefont {Zhu}}, \bibinfo {author} {\bibfnamefont
  {K.}~\bibnamefont {Yin}}, \bibinfo {author} {\bibfnamefont {T.}~\bibnamefont
  {Li}}, \bibinfo {author} {\bibfnamefont {D.}~\bibnamefont {Wu}},\ and\
  \bibinfo {author} {\bibfnamefont {T.}~\bibnamefont {Jiang}},\ }\bibfield
  {title} {\bibinfo {title} {Low-latency deep-reinforcement learning algorithm
  for ultrafast fiber lasers},\ }\href {https://doi.org/10.1364/PRJ.428117}
  {\bibfield  {journal} {\bibinfo  {journal} {Photonics Research}\ }\textbf
  {\bibinfo {volume} {9}},\ \bibinfo {pages} {1493} (\bibinfo {year}
  {2021})}\BibitemShut {NoStop}%
\bibitem [{\citenamefont {Muiños-Landin}\ \emph {et~al.}(2021)\citenamefont
  {Muiños-Landin}, \citenamefont {Fischer}, \citenamefont {Holubec},\ and\
  \citenamefont {Cichos}}]{MuinosLandin2021}%
  \BibitemOpen
  \bibfield  {author} {\bibinfo {author} {\bibfnamefont {S.}~\bibnamefont
  {Muiños-Landin}}, \bibinfo {author} {\bibfnamefont {A.}~\bibnamefont
  {Fischer}}, \bibinfo {author} {\bibfnamefont {V.}~\bibnamefont {Holubec}},\
  and\ \bibinfo {author} {\bibfnamefont {F.}~\bibnamefont {Cichos}},\
  }\bibfield  {title} {\bibinfo {title} {Reinforcement learning with artificial
  microswimmers},\ }\href {https://doi.org/10.1126/scirobotics.abd9285}
  {\bibfield  {journal} {\bibinfo  {journal} {Science Robotics}\ }\textbf
  {\bibinfo {volume} {6}},\ \bibinfo {pages} {eabd9285} (\bibinfo {year}
  {2021})}\BibitemShut {NoStop}%
\bibitem [{\citenamefont {Baum}\ \emph {et~al.}(2021)\citenamefont {Baum},
  \citenamefont {Amico}, \citenamefont {Howell}, \citenamefont {Hush},
  \citenamefont {Liuzzi}, \citenamefont {Mundada}, \citenamefont {Merkh},
  \citenamefont {Carvalho},\ and\ \citenamefont {Biercuk}}]{Baum2021a}%
  \BibitemOpen
  \bibfield  {author} {\bibinfo {author} {\bibfnamefont {Y.}~\bibnamefont
  {Baum}}, \bibinfo {author} {\bibfnamefont {M.}~\bibnamefont {Amico}},
  \bibinfo {author} {\bibfnamefont {S.}~\bibnamefont {Howell}}, \bibinfo
  {author} {\bibfnamefont {M.}~\bibnamefont {Hush}}, \bibinfo {author}
  {\bibfnamefont {M.}~\bibnamefont {Liuzzi}}, \bibinfo {author} {\bibfnamefont
  {P.}~\bibnamefont {Mundada}}, \bibinfo {author} {\bibfnamefont
  {T.}~\bibnamefont {Merkh}}, \bibinfo {author} {\bibfnamefont {A.~R.}\
  \bibnamefont {Carvalho}},\ and\ \bibinfo {author} {\bibfnamefont {M.~J.}\
  \bibnamefont {Biercuk}},\ }\bibfield  {title} {\bibinfo {title} {Experimental
  deep reinforcement learning for error-robust gate-set design on a
  superconducting quantum computer},\ }\href
  {https://doi.org/10.1103/PRXQuantum.2.040324} {\bibfield  {journal} {\bibinfo
   {journal} {PRX Quantum}\ }\textbf {\bibinfo {volume} {2}},\ \bibinfo {pages}
  {040324} (\bibinfo {year} {2021})}\BibitemShut {NoStop}%
\bibitem [{\citenamefont {Nguyen}\ \emph {et~al.}(2021)\citenamefont {Nguyen},
  \citenamefont {Orbell}, \citenamefont {Lennon}, \citenamefont {Moon},
  \citenamefont {Vigneau}, \citenamefont {Camenzind}, \citenamefont {Yu},
  \citenamefont {Zumb{\"u}hl}, \citenamefont {Briggs}, \citenamefont {Osborne},
  \citenamefont {Sejdinovic},\ and\ \citenamefont {Ares}}]{Nguyen2021c}%
  \BibitemOpen
  \bibfield  {author} {\bibinfo {author} {\bibfnamefont {V.}~\bibnamefont
  {Nguyen}}, \bibinfo {author} {\bibfnamefont {S.~B.}\ \bibnamefont {Orbell}},
  \bibinfo {author} {\bibfnamefont {D.~T.}\ \bibnamefont {Lennon}}, \bibinfo
  {author} {\bibfnamefont {H.}~\bibnamefont {Moon}}, \bibinfo {author}
  {\bibfnamefont {F.}~\bibnamefont {Vigneau}}, \bibinfo {author} {\bibfnamefont
  {L.~C.}\ \bibnamefont {Camenzind}}, \bibinfo {author} {\bibfnamefont
  {L.}~\bibnamefont {Yu}}, \bibinfo {author} {\bibfnamefont {D.~M.}\
  \bibnamefont {Zumb{\"u}hl}}, \bibinfo {author} {\bibfnamefont {G.~A.~D.}\
  \bibnamefont {Briggs}}, \bibinfo {author} {\bibfnamefont {M.~A.}\
  \bibnamefont {Osborne}}, \bibinfo {author} {\bibfnamefont {D.}~\bibnamefont
  {Sejdinovic}},\ and\ \bibinfo {author} {\bibfnamefont {N.}~\bibnamefont
  {Ares}},\ }\bibfield  {title} {\bibinfo {title} {Deep reinforcement learning
  for efficient measurement of quantum devices},\ }\href
  {https://doi.org/10.1038/s41534-021-00434-x} {\bibfield  {journal} {\bibinfo
  {journal} {npj Quantum Information}\ }\textbf {\bibinfo {volume} {7}},\
  \bibinfo {pages} {100} (\bibinfo {year} {2021})}\BibitemShut {NoStop}%
\bibitem [{\citenamefont {Li}\ \emph {et~al.}(2022)\citenamefont {Li},
  \citenamefont {Yang}, \citenamefont {Xiao}, \citenamefont {Zhang},
  \citenamefont {Li}, \citenamefont {Han}, \citenamefont {Liu}, \citenamefont
  {Ouyang},\ and\ \citenamefont {Zhu}}]{li2022deep}%
  \BibitemOpen
  \bibfield  {author} {\bibinfo {author} {\bibfnamefont {Z.}~\bibnamefont
  {Li}}, \bibinfo {author} {\bibfnamefont {S.}~\bibnamefont {Yang}}, \bibinfo
  {author} {\bibfnamefont {Q.}~\bibnamefont {Xiao}}, \bibinfo {author}
  {\bibfnamefont {T.}~\bibnamefont {Zhang}}, \bibinfo {author} {\bibfnamefont
  {Y.}~\bibnamefont {Li}}, \bibinfo {author} {\bibfnamefont {L.}~\bibnamefont
  {Han}}, \bibinfo {author} {\bibfnamefont {D.}~\bibnamefont {Liu}}, \bibinfo
  {author} {\bibfnamefont {X.}~\bibnamefont {Ouyang}},\ and\ \bibinfo {author}
  {\bibfnamefont {J.}~\bibnamefont {Zhu}},\ }\bibfield  {title} {\bibinfo
  {title} {{Deep reinforcement with spectrum series learning control for a
  mode-locked fiber laser}},\ }\href {https://doi.org/10.1364/PRJ.455493}
  {\bibfield  {journal} {\bibinfo  {journal} {Photonics Research}\ }\textbf
  {\bibinfo {volume} {10}},\ \bibinfo {pages} {1491} (\bibinfo {year}
  {2022})}\BibitemShut {NoStop}%
\bibitem [{\citenamefont {Chen}\ \emph {et~al.}(2013)\citenamefont {Chen},
  \citenamefont {Dong}, \citenamefont {Li}, \citenamefont {Chu},\ and\
  \citenamefont {Tarn}}]{chen2013fidelity}%
  \BibitemOpen
  \bibfield  {author} {\bibinfo {author} {\bibfnamefont {C.}~\bibnamefont
  {Chen}}, \bibinfo {author} {\bibfnamefont {D.}~\bibnamefont {Dong}}, \bibinfo
  {author} {\bibfnamefont {H.-X.}\ \bibnamefont {Li}}, \bibinfo {author}
  {\bibfnamefont {J.}~\bibnamefont {Chu}},\ and\ \bibinfo {author}
  {\bibfnamefont {T.-J.}\ \bibnamefont {Tarn}},\ }\bibfield  {title} {\bibinfo
  {title} {{Fidelity-Based Probabilistic Q-Learning for Control of Quantum
  Systems}},\ }\href {http://doi.org/10.1109/TNNLS.2013.2283574} {\bibfield
  {journal} {\bibinfo  {journal} {IEEE Transactions on Neural Networks and
  Learning Systems}\ }\textbf {\bibinfo {volume} {25}},\ \bibinfo {pages} {920}
  (\bibinfo {year} {2013})}\BibitemShut {NoStop}%
\bibitem [{\citenamefont {Bukov}\ \emph {et~al.}(2018)\citenamefont {Bukov},
  \citenamefont {Day}, \citenamefont {Sels}, \citenamefont {Weinberg},
  \citenamefont {Polkovnikov},\ and\ \citenamefont {Mehta}}]{Bukov2018}%
  \BibitemOpen
  \bibfield  {author} {\bibinfo {author} {\bibfnamefont {M.}~\bibnamefont
  {Bukov}}, \bibinfo {author} {\bibfnamefont {A.~G.~R.}\ \bibnamefont {Day}},
  \bibinfo {author} {\bibfnamefont {D.}~\bibnamefont {Sels}}, \bibinfo {author}
  {\bibfnamefont {P.}~\bibnamefont {Weinberg}}, \bibinfo {author}
  {\bibfnamefont {A.}~\bibnamefont {Polkovnikov}},\ and\ \bibinfo {author}
  {\bibfnamefont {P.}~\bibnamefont {Mehta}},\ }\bibfield  {title} {\bibinfo
  {title} {Reinforcement learning in different phases of quantum control},\
  }\href {https://doi.org/10.1103/PhysRevX.8.031086} {\bibfield  {journal}
  {\bibinfo  {journal} {Phys. Rev. X}\ }\textbf {\bibinfo {volume} {8}},\
  \bibinfo {pages} {031086} (\bibinfo {year} {2018})}\BibitemShut {NoStop}%
\bibitem [{\citenamefont {Borah}\ \emph {et~al.}(2021)\citenamefont {Borah},
  \citenamefont {Sarma}, \citenamefont {Kewming}, \citenamefont {Milburn},\
  and\ \citenamefont {Twamley}}]{borah2021measurement}%
  \BibitemOpen
  \bibfield  {author} {\bibinfo {author} {\bibfnamefont {S.}~\bibnamefont
  {Borah}}, \bibinfo {author} {\bibfnamefont {B.}~\bibnamefont {Sarma}},
  \bibinfo {author} {\bibfnamefont {M.}~\bibnamefont {Kewming}}, \bibinfo
  {author} {\bibfnamefont {G.~J.}\ \bibnamefont {Milburn}},\ and\ \bibinfo
  {author} {\bibfnamefont {J.}~\bibnamefont {Twamley}},\ }\bibfield  {title}
  {\bibinfo {title} {{Measurement-Based Feedback Quantum Control with Deep
  Reinforcement Learning for a Double-Well Nonlinear Potential}},\ }\href
  {https://doi.org/10.1103/PhysRevLett.127.190403} {\bibfield  {journal}
  {\bibinfo  {journal} {Physical Review Letters}\ }\textbf {\bibinfo {volume}
  {127}},\ \bibinfo {pages} {190403} (\bibinfo {year} {2021})}\BibitemShut
  {NoStop}%
\bibitem [{\citenamefont {Sivak}\ \emph {et~al.}(2022)\citenamefont {Sivak},
  \citenamefont {Eickbusch}, \citenamefont {Liu}, \citenamefont {Royer},
  \citenamefont {Tsioutsios},\ and\ \citenamefont {Devoret}}]{sivak2021model}%
  \BibitemOpen
  \bibfield  {author} {\bibinfo {author} {\bibfnamefont {V.~V.}\ \bibnamefont
  {Sivak}}, \bibinfo {author} {\bibfnamefont {A.}~\bibnamefont {Eickbusch}},
  \bibinfo {author} {\bibfnamefont {H.}~\bibnamefont {Liu}}, \bibinfo {author}
  {\bibfnamefont {B.}~\bibnamefont {Royer}}, \bibinfo {author} {\bibfnamefont
  {I.}~\bibnamefont {Tsioutsios}},\ and\ \bibinfo {author} {\bibfnamefont
  {M.~H.}\ \bibnamefont {Devoret}},\ }\bibfield  {title} {\bibinfo {title}
  {Model-{F}ree {Q}uantum {C}ontrol with {R}einforcement {L}earning},\ }\href
  {https://doi.org/10.1103/PhysRevX.12.011059} {\bibfield  {journal} {\bibinfo
  {journal} {Phys. Rev. X}\ }\textbf {\bibinfo {volume} {12}},\ \bibinfo
  {pages} {011059} (\bibinfo {year} {2022})}\BibitemShut {NoStop}%
\bibitem [{\citenamefont {Porotti}\ \emph {et~al.}(2022)\citenamefont
  {Porotti}, \citenamefont {Essig}, \citenamefont {Huard},\ and\ \citenamefont
  {Marquardt}}]{porotti2022deep}%
  \BibitemOpen
  \bibfield  {author} {\bibinfo {author} {\bibfnamefont {R.}~\bibnamefont
  {Porotti}}, \bibinfo {author} {\bibfnamefont {A.}~\bibnamefont {Essig}},
  \bibinfo {author} {\bibfnamefont {B.}~\bibnamefont {Huard}},\ and\ \bibinfo
  {author} {\bibfnamefont {F.}~\bibnamefont {Marquardt}},\ }\bibfield  {title}
  {\bibinfo {title} {{Deep Reinforcement Learning for Quantum State Preparation
  with Weak Nonlinear Measurements}},\ }\href
  {https://doi.org/10.22331/q-2022-06-28-747} {\bibfield  {journal} {\bibinfo
  {journal} {Quantum}\ }\textbf {\bibinfo {volume} {6}},\ \bibinfo {pages}
  {747} (\bibinfo {year} {2022})}\BibitemShut {NoStop}%
\bibitem [{\citenamefont {Niu}\ \emph {et~al.}(2019)\citenamefont {Niu},
  \citenamefont {Boixo}, \citenamefont {Smelyanskiy},\ and\ \citenamefont
  {Neven}}]{niu2019universal}%
  \BibitemOpen
  \bibfield  {author} {\bibinfo {author} {\bibfnamefont {M.~Y.}\ \bibnamefont
  {Niu}}, \bibinfo {author} {\bibfnamefont {S.}~\bibnamefont {Boixo}}, \bibinfo
  {author} {\bibfnamefont {V.~N.}\ \bibnamefont {Smelyanskiy}},\ and\ \bibinfo
  {author} {\bibfnamefont {H.}~\bibnamefont {Neven}},\ }\bibfield  {title}
  {\bibinfo {title} {Universal quantum control through deep reinforcement
  learning},\ }\href {https://doi.org/10.1038/s41534-019-0141-3} {\bibfield
  {journal} {\bibinfo  {journal} {npj Quantum Information}\ }\textbf {\bibinfo
  {volume} {5}},\ \bibinfo {pages} {1} (\bibinfo {year} {2019})}\BibitemShut
  {NoStop}%
\bibitem [{\citenamefont {F\"osel}\ \emph {et~al.}(2018)\citenamefont
  {F\"osel}, \citenamefont {Tighineanu}, \citenamefont {Weiss},\ and\
  \citenamefont {Marquardt}}]{Fosel2018}%
  \BibitemOpen
  \bibfield  {author} {\bibinfo {author} {\bibfnamefont {T.}~\bibnamefont
  {F\"osel}}, \bibinfo {author} {\bibfnamefont {P.}~\bibnamefont {Tighineanu}},
  \bibinfo {author} {\bibfnamefont {T.}~\bibnamefont {Weiss}},\ and\ \bibinfo
  {author} {\bibfnamefont {F.}~\bibnamefont {Marquardt}},\ }\bibfield  {title}
  {\bibinfo {title} {Reinforcement learning with neural networks for quantum
  feedback},\ }\href {https://doi.org/10.1103/PhysRevX.8.031084} {\bibfield
  {journal} {\bibinfo  {journal} {Phys. Rev. X}\ }\textbf {\bibinfo {volume}
  {8}},\ \bibinfo {pages} {031084} (\bibinfo {year} {2018})}\BibitemShut
  {NoStop}%
\bibitem [{\citenamefont {Nautrup}\ \emph {et~al.}(2019)\citenamefont
  {Nautrup}, \citenamefont {Delfosse}, \citenamefont {Dunjko}, \citenamefont
  {Briegel},\ and\ \citenamefont {Friis}}]{nautrup2019optimizing}%
  \BibitemOpen
  \bibfield  {author} {\bibinfo {author} {\bibfnamefont {H.~P.}\ \bibnamefont
  {Nautrup}}, \bibinfo {author} {\bibfnamefont {N.}~\bibnamefont {Delfosse}},
  \bibinfo {author} {\bibfnamefont {V.}~\bibnamefont {Dunjko}}, \bibinfo
  {author} {\bibfnamefont {H.~J.}\ \bibnamefont {Briegel}},\ and\ \bibinfo
  {author} {\bibfnamefont {N.}~\bibnamefont {Friis}},\ }\bibfield  {title}
  {\bibinfo {title} {{Optimizing Quantum Error Correction Codes with
  Reinforcement Learning}},\ }\href {https://doi.org/10.22331/q-2019-12-16-215}
  {\bibfield  {journal} {\bibinfo  {journal} {Quantum}\ }\textbf {\bibinfo
  {volume} {3}},\ \bibinfo {pages} {215} (\bibinfo {year} {2019})}\BibitemShut
  {NoStop}%
\bibitem [{\citenamefont {Sweke}\ \emph {et~al.}(2020)\citenamefont {Sweke},
  \citenamefont {Kesselring}, \citenamefont {van Nieuwenburg},\ and\
  \citenamefont {Eisert}}]{sweke2020reinforcement}%
  \BibitemOpen
  \bibfield  {author} {\bibinfo {author} {\bibfnamefont {R.}~\bibnamefont
  {Sweke}}, \bibinfo {author} {\bibfnamefont {M.~S.}\ \bibnamefont
  {Kesselring}}, \bibinfo {author} {\bibfnamefont {E.~P.}\ \bibnamefont {van
  Nieuwenburg}},\ and\ \bibinfo {author} {\bibfnamefont {J.}~\bibnamefont
  {Eisert}},\ }\bibfield  {title} {\bibinfo {title} {Reinforcement learning
  decoders for fault-tolerant quantum computation},\ }\href
  {https://doi.org/10.1088/2632-2153/abc609} {\bibfield  {journal} {\bibinfo
  {journal} {Machine Learning: Science and Technology}\ }\textbf {\bibinfo
  {volume} {2}},\ \bibinfo {pages} {025005} (\bibinfo {year}
  {2020})}\BibitemShut {NoStop}%
\bibitem [{\citenamefont {Zhang}\ \emph {et~al.}(2020)\citenamefont {Zhang},
  \citenamefont {Zheng}, \citenamefont {Zhang},\ and\ \citenamefont
  {Deng}}]{zhang2020topological}%
  \BibitemOpen
  \bibfield  {author} {\bibinfo {author} {\bibfnamefont {Y.-H.}\ \bibnamefont
  {Zhang}}, \bibinfo {author} {\bibfnamefont {P.-L.}\ \bibnamefont {Zheng}},
  \bibinfo {author} {\bibfnamefont {Y.}~\bibnamefont {Zhang}},\ and\ \bibinfo
  {author} {\bibfnamefont {D.-L.}\ \bibnamefont {Deng}},\ }\bibfield  {title}
  {\bibinfo {title} {{Topological Quantum Compiling with Reinforcement
  Learning}},\ }\href {https://doi.org/10.1103/PhysRevLett.125.170501}
  {\bibfield  {journal} {\bibinfo  {journal} {Physical Review Letters}\
  }\textbf {\bibinfo {volume} {125}},\ \bibinfo {pages} {170501} (\bibinfo
  {year} {2020})}\BibitemShut {NoStop}%
\bibitem [{\citenamefont {F{\"o}sel}\ \emph {et~al.}(2021)\citenamefont
  {F{\"o}sel}, \citenamefont {Niu}, \citenamefont {Marquardt},\ and\
  \citenamefont {Li}}]{fosel2021quantum}%
  \BibitemOpen
  \bibfield  {author} {\bibinfo {author} {\bibfnamefont {T.}~\bibnamefont
  {F{\"o}sel}}, \bibinfo {author} {\bibfnamefont {M.~Y.}\ \bibnamefont {Niu}},
  \bibinfo {author} {\bibfnamefont {F.}~\bibnamefont {Marquardt}},\ and\
  \bibinfo {author} {\bibfnamefont {L.}~\bibnamefont {Li}},\ }\bibfield
  {title} {\bibinfo {title} {Quantum circuit optimization with deep
  reinforcement learning},\ }\href {https://doi.org/10.48550/arXiv.2103.07585}
  {\bibfield  {journal} {\bibinfo  {journal} {arXiv:2103.07585}\ } (\bibinfo
  {year} {2021})}\BibitemShut {NoStop}%
\bibitem [{\citenamefont {Carleo}\ \emph {et~al.}(2019)\citenamefont {Carleo},
  \citenamefont {Cirac}, \citenamefont {Cranmer}, \citenamefont {Daudet},
  \citenamefont {Schuld}, \citenamefont {Tishby}, \citenamefont
  {Vogt-Maranto},\ and\ \citenamefont {Zdeborová}}]{Carleo2019}%
  \BibitemOpen
  \bibfield  {author} {\bibinfo {author} {\bibfnamefont {G.}~\bibnamefont
  {Carleo}}, \bibinfo {author} {\bibfnamefont {I.}~\bibnamefont {Cirac}},
  \bibinfo {author} {\bibfnamefont {K.}~\bibnamefont {Cranmer}}, \bibinfo
  {author} {\bibfnamefont {L.}~\bibnamefont {Daudet}}, \bibinfo {author}
  {\bibfnamefont {M.}~\bibnamefont {Schuld}}, \bibinfo {author} {\bibfnamefont
  {N.}~\bibnamefont {Tishby}}, \bibinfo {author} {\bibfnamefont
  {L.}~\bibnamefont {Vogt-Maranto}},\ and\ \bibinfo {author} {\bibfnamefont
  {L.}~\bibnamefont {Zdeborová}},\ }\bibfield  {title} {\bibinfo {title}
  {Machine learning and the physical sciences},\ }\href
  {https://doi.org/10.1103/RevModPhys.91.045002} {\bibfield  {journal}
  {\bibinfo  {journal} {Review of Modern Physics}\ }\textbf {\bibinfo {volume}
  {91}},\ \bibinfo {pages} {045002} (\bibinfo {year} {2019})}\BibitemShut
  {NoStop}%
\bibitem [{\citenamefont {Dawid}\ \emph {et~al.}(2022)\citenamefont {Dawid},
  \citenamefont {Arnold}, \citenamefont {Requena}, \citenamefont {Gresch},
  \citenamefont {Płodzień}, \citenamefont {Donatella}, \citenamefont
  {Nicoli}, \citenamefont {Stornati}, \citenamefont {Koch}, \citenamefont
  {Büttner}, \citenamefont {Okuła}, \citenamefont {Muñoz-Gil}, \citenamefont
  {Vargas-Hernández}, \citenamefont {Cervera-Lierta}, \citenamefont
  {Carrasquilla}, \citenamefont {Dunjko}, \citenamefont {Gabrié},
  \citenamefont {Huembeli}, \citenamefont {van Nieuwenburg}, \citenamefont
  {Vicentini}, \citenamefont {Wang}, \citenamefont {Wetzel}, \citenamefont
  {Carleo}, \citenamefont {Greplová}, \citenamefont {Krems}, \citenamefont
  {Marquardt}, \citenamefont {Tomza}, \citenamefont {Lewenstein},\ and\
  \citenamefont {Dauphin}}]{Dawid2022}%
  \BibitemOpen
  \bibfield  {author} {\bibinfo {author} {\bibfnamefont {A.}~\bibnamefont
  {Dawid}}, \bibinfo {author} {\bibfnamefont {J.}~\bibnamefont {Arnold}},
  \bibinfo {author} {\bibfnamefont {B.}~\bibnamefont {Requena}}, \bibinfo
  {author} {\bibfnamefont {A.}~\bibnamefont {Gresch}}, \bibinfo {author}
  {\bibfnamefont {M.}~\bibnamefont {Płodzień}}, \bibinfo {author}
  {\bibfnamefont {K.}~\bibnamefont {Donatella}}, \bibinfo {author}
  {\bibfnamefont {K.~A.}\ \bibnamefont {Nicoli}}, \bibinfo {author}
  {\bibfnamefont {P.}~\bibnamefont {Stornati}}, \bibinfo {author}
  {\bibfnamefont {R.}~\bibnamefont {Koch}}, \bibinfo {author} {\bibfnamefont
  {M.}~\bibnamefont {Büttner}}, \bibinfo {author} {\bibfnamefont
  {R.}~\bibnamefont {Okuła}}, \bibinfo {author} {\bibfnamefont
  {G.}~\bibnamefont {Muñoz-Gil}}, \bibinfo {author} {\bibfnamefont {R.~A.}\
  \bibnamefont {Vargas-Hernández}}, \bibinfo {author} {\bibfnamefont
  {A.}~\bibnamefont {Cervera-Lierta}}, \bibinfo {author} {\bibfnamefont
  {J.}~\bibnamefont {Carrasquilla}}, \bibinfo {author} {\bibfnamefont
  {V.}~\bibnamefont {Dunjko}}, \bibinfo {author} {\bibfnamefont
  {M.}~\bibnamefont {Gabrié}}, \bibinfo {author} {\bibfnamefont
  {P.}~\bibnamefont {Huembeli}}, \bibinfo {author} {\bibfnamefont
  {E.}~\bibnamefont {van Nieuwenburg}}, \bibinfo {author} {\bibfnamefont
  {F.}~\bibnamefont {Vicentini}}, \bibinfo {author} {\bibfnamefont
  {L.}~\bibnamefont {Wang}}, \bibinfo {author} {\bibfnamefont {S.~J.}\
  \bibnamefont {Wetzel}}, \bibinfo {author} {\bibfnamefont {G.}~\bibnamefont
  {Carleo}}, \bibinfo {author} {\bibfnamefont {E.}~\bibnamefont {Greplová}},
  \bibinfo {author} {\bibfnamefont {R.}~\bibnamefont {Krems}}, \bibinfo
  {author} {\bibfnamefont {F.}~\bibnamefont {Marquardt}}, \bibinfo {author}
  {\bibfnamefont {M.}~\bibnamefont {Tomza}}, \bibinfo {author} {\bibfnamefont
  {M.}~\bibnamefont {Lewenstein}},\ and\ \bibinfo {author} {\bibfnamefont
  {A.}~\bibnamefont {Dauphin}},\ }\bibfield  {title} {\bibinfo {title} {Modern
  applications of machine learning in quantum sciences},\ }\href@noop {}
  {\bibfield  {journal} {\bibinfo  {journal} {arXiv:2204.04198}\ } (\bibinfo
  {year} {2022})},\ \Eprint {https://arxiv.org/abs/2204.04198}
  {arXiv:2204.04198 [quant-ph]} \BibitemShut {NoStop}%
\bibitem [{\citenamefont {Krenn}\ \emph {et~al.}(2022)\citenamefont {Krenn},
  \citenamefont {Landgraf}, \citenamefont {Foesel},\ and\ \citenamefont
  {Marquardt}}]{Krenn2022}%
  \BibitemOpen
  \bibfield  {author} {\bibinfo {author} {\bibfnamefont {M.}~\bibnamefont
  {Krenn}}, \bibinfo {author} {\bibfnamefont {J.}~\bibnamefont {Landgraf}},
  \bibinfo {author} {\bibfnamefont {T.}~\bibnamefont {Foesel}},\ and\ \bibinfo
  {author} {\bibfnamefont {F.}~\bibnamefont {Marquardt}},\ }\bibfield  {title}
  {\bibinfo {title} {Artificial intelligence and machine learning for quantum
  technologies},\ }\href {https://arxiv.org/abs/2208.03836} {\bibfield
  {journal} {\bibinfo  {journal} {arXiv:2208.03836}\ } (\bibinfo {year}
  {2022})}\BibitemShut {NoStop}%
\bibitem [{\citenamefont {Blais}\ \emph {et~al.}(2004)\citenamefont {Blais},
  \citenamefont {Huang}, \citenamefont {Wallraff}, \citenamefont {Girvin},\
  and\ \citenamefont {Schoelkopf}}]{Blais2004}%
  \BibitemOpen
  \bibfield  {author} {\bibinfo {author} {\bibfnamefont {A.}~\bibnamefont
  {Blais}}, \bibinfo {author} {\bibfnamefont {R.-S.}\ \bibnamefont {Huang}},
  \bibinfo {author} {\bibfnamefont {A.}~\bibnamefont {Wallraff}}, \bibinfo
  {author} {\bibfnamefont {S.~M.}\ \bibnamefont {Girvin}},\ and\ \bibinfo
  {author} {\bibfnamefont {R.~J.}\ \bibnamefont {Schoelkopf}},\ }\bibfield
  {title} {\bibinfo {title} {Cavity quantum electrodynamics for superconducting
  electrical circuits: An architecture for quantum computation},\ }\href
  {https://doi.org/10.1103/PhysRevA.69.062320} {\bibfield  {journal} {\bibinfo
  {journal} {Phys. Rev. A}\ }\textbf {\bibinfo {volume} {69}},\ \bibinfo
  {pages} {062320} (\bibinfo {year} {2004})}\BibitemShut {NoStop}%
\bibitem [{\citenamefont {Wallraff}\ \emph {et~al.}(2005)\citenamefont
  {Wallraff}, \citenamefont {Schuster}, \citenamefont {Blais}, \citenamefont
  {Frunzio}, \citenamefont {Majer}, \citenamefont {Devoret}, \citenamefont
  {Girvin},\ and\ \citenamefont {Schoelkopf}}]{Wallraff2005}%
  \BibitemOpen
  \bibfield  {author} {\bibinfo {author} {\bibfnamefont {A.}~\bibnamefont
  {Wallraff}}, \bibinfo {author} {\bibfnamefont {D.~I.}\ \bibnamefont
  {Schuster}}, \bibinfo {author} {\bibfnamefont {A.}~\bibnamefont {Blais}},
  \bibinfo {author} {\bibfnamefont {L.}~\bibnamefont {Frunzio}}, \bibinfo
  {author} {\bibfnamefont {J.}~\bibnamefont {Majer}}, \bibinfo {author}
  {\bibfnamefont {M.~H.}\ \bibnamefont {Devoret}}, \bibinfo {author}
  {\bibfnamefont {S.~M.}\ \bibnamefont {Girvin}},\ and\ \bibinfo {author}
  {\bibfnamefont {R.~J.}\ \bibnamefont {Schoelkopf}},\ }\bibfield  {title}
  {\bibinfo {title} {Approaching unit visibility for control of a
  superconducting qubit with dispersive readout},\ }\href
  {https://doi.org/10.1103/PhysRevLett.95.060501} {\bibfield  {journal}
  {\bibinfo  {journal} {Phys. Rev. Lett.}\ }\textbf {\bibinfo {volume} {95}},\
  \bibinfo {pages} {060501} (\bibinfo {year} {2005})}\BibitemShut {NoStop}%
\bibitem [{\citenamefont {Gambetta}\ \emph {et~al.}(2007)\citenamefont
  {Gambetta}, \citenamefont {Braff}, \citenamefont {Wallraff}, \citenamefont
  {Girvin},\ and\ \citenamefont {Schoelkopf}}]{Gambetta2007}%
  \BibitemOpen
  \bibfield  {author} {\bibinfo {author} {\bibfnamefont {J.}~\bibnamefont
  {Gambetta}}, \bibinfo {author} {\bibfnamefont {W.~A.}\ \bibnamefont {Braff}},
  \bibinfo {author} {\bibfnamefont {A.}~\bibnamefont {Wallraff}}, \bibinfo
  {author} {\bibfnamefont {S.~M.}\ \bibnamefont {Girvin}},\ and\ \bibinfo
  {author} {\bibfnamefont {R.~J.}\ \bibnamefont {Schoelkopf}},\ }\bibfield
  {title} {\bibinfo {title} {Protocols for optimal readout of qubits using a
  continuous quantum nondemolition measurement},\ }\href
  {https://doi.org/10.1103/PhysRevA.76.012325} {\bibfield  {journal} {\bibinfo
  {journal} {Phys. Rev. A}\ }\textbf {\bibinfo {volume} {76}},\ \bibinfo
  {pages} {012325} (\bibinfo {year} {2007})}\BibitemShut {NoStop}%
\bibitem [{\citenamefont {Goodfellow}\ \emph {et~al.}(2016)\citenamefont
  {Goodfellow}, \citenamefont {Bengio},\ and\ \citenamefont
  {Courville}}]{Goodfellow2016}%
  \BibitemOpen
  \bibfield  {author} {\bibinfo {author} {\bibfnamefont {I.}~\bibnamefont
  {Goodfellow}}, \bibinfo {author} {\bibfnamefont {Y.}~\bibnamefont {Bengio}},\
  and\ \bibinfo {author} {\bibfnamefont {A.}~\bibnamefont {Courville}},\
  }\href@noop {} {\emph {\bibinfo {title} {Deep Learning}}}\ (\bibinfo
  {publisher} {MIT Press},\ \bibinfo {year} {2016})\ \bibinfo {note}
  {\url{http://www.deeplearningbook.org}}\BibitemShut {NoStop}%
\bibitem [{\citenamefont {Hochreiter}\ and\ \citenamefont
  {Schmidhuber}(1997)}]{Hochreiter1997}%
  \BibitemOpen
  \bibfield  {author} {\bibinfo {author} {\bibfnamefont {S.}~\bibnamefont
  {Hochreiter}}\ and\ \bibinfo {author} {\bibfnamefont {J.}~\bibnamefont
  {Schmidhuber}},\ }\bibfield  {title} {\bibinfo {title} {Long short-term
  memory},\ }\href {https://doi.org/10.1162/neco.1997.9.8.1735} {\bibfield
  {journal} {\bibinfo  {journal} {Neural Computation}\ }\textbf {\bibinfo
  {volume} {9}},\ \bibinfo {pages} {1735} (\bibinfo {year} {1997})}\BibitemShut
  {NoStop}%
\bibitem [{\citenamefont {Schulman}\ \emph {et~al.}(2017)\citenamefont
  {Schulman}, \citenamefont {Wolski}, \citenamefont {Dhariwal}, \citenamefont
  {Radford},\ and\ \citenamefont {Klimov}}]{schulman_proximal_2017}%
  \BibitemOpen
  \bibfield  {author} {\bibinfo {author} {\bibfnamefont {J.}~\bibnamefont
  {Schulman}}, \bibinfo {author} {\bibfnamefont {F.}~\bibnamefont {Wolski}},
  \bibinfo {author} {\bibfnamefont {P.}~\bibnamefont {Dhariwal}}, \bibinfo
  {author} {\bibfnamefont {A.}~\bibnamefont {Radford}},\ and\ \bibinfo {author}
  {\bibfnamefont {O.}~\bibnamefont {Klimov}},\ }\bibfield  {title} {\bibinfo
  {title} {Proximal {Policy} {Optimization} {Algorithms}},\ }\href
  {http://arxiv.org/abs/1707.06347} {\bibfield  {journal} {\bibinfo  {journal}
  {arXiv:1707.06347 [cs]}\ } (\bibinfo {year} {2017})}\BibitemShut {NoStop}%
\bibitem [{\citenamefont {Hill}\ \emph {et~al.}(2018)\citenamefont {Hill},
  \citenamefont {Raffin}, \citenamefont {Ernestus}, \citenamefont {Gleave},
  \citenamefont {Kanervisto}, \citenamefont {Traore}, \citenamefont {Dhariwal},
  \citenamefont {Hesse}, \citenamefont {Klimov}, \citenamefont {Nichol},
  \citenamefont {Plappert}, \citenamefont {Radford}, \citenamefont {Schulman},
  \citenamefont {Sidor},\ and\ \citenamefont {Wu}}]{hill_stable_2018}%
  \BibitemOpen
  \bibfield  {author} {\bibinfo {author} {\bibfnamefont {A.}~\bibnamefont
  {Hill}}, \bibinfo {author} {\bibfnamefont {A.}~\bibnamefont {Raffin}},
  \bibinfo {author} {\bibfnamefont {M.}~\bibnamefont {Ernestus}}, \bibinfo
  {author} {\bibfnamefont {A.}~\bibnamefont {Gleave}}, \bibinfo {author}
  {\bibfnamefont {A.}~\bibnamefont {Kanervisto}}, \bibinfo {author}
  {\bibfnamefont {R.}~\bibnamefont {Traore}}, \bibinfo {author} {\bibfnamefont
  {P.}~\bibnamefont {Dhariwal}}, \bibinfo {author} {\bibfnamefont
  {C.}~\bibnamefont {Hesse}}, \bibinfo {author} {\bibfnamefont
  {O.}~\bibnamefont {Klimov}}, \bibinfo {author} {\bibfnamefont
  {A.}~\bibnamefont {Nichol}}, \bibinfo {author} {\bibfnamefont
  {M.}~\bibnamefont {Plappert}}, \bibinfo {author} {\bibfnamefont
  {A.}~\bibnamefont {Radford}}, \bibinfo {author} {\bibfnamefont
  {J.}~\bibnamefont {Schulman}}, \bibinfo {author} {\bibfnamefont
  {S.}~\bibnamefont {Sidor}},\ and\ \bibinfo {author} {\bibfnamefont
  {Y.}~\bibnamefont {Wu}},\ }\href@noop {} {\bibinfo {title} {Stable
  baselines}},\ \bibinfo {howpublished}
  {\url{https://github.com/hill-a/stable-baselines}} (\bibinfo {year}
  {2018})\BibitemShut {NoStop}%
\bibitem [{\citenamefont {Magnard}\ \emph {et~al.}(2018)\citenamefont
  {Magnard}, \citenamefont {Kurpiers}, \citenamefont {Royer}, \citenamefont
  {Walter}, \citenamefont {Besse}, \citenamefont {Gasparinetti}, \citenamefont
  {Pechal}, \citenamefont {Heinsoo}, \citenamefont {Storz}, \citenamefont
  {Blais},\ and\ \citenamefont {Wallraff}}]{Magnard2018}%
  \BibitemOpen
  \bibfield  {author} {\bibinfo {author} {\bibfnamefont {P.}~\bibnamefont
  {Magnard}}, \bibinfo {author} {\bibfnamefont {P.}~\bibnamefont {Kurpiers}},
  \bibinfo {author} {\bibfnamefont {B.}~\bibnamefont {Royer}}, \bibinfo
  {author} {\bibfnamefont {T.}~\bibnamefont {Walter}}, \bibinfo {author}
  {\bibfnamefont {J.-C.}\ \bibnamefont {Besse}}, \bibinfo {author}
  {\bibfnamefont {S.}~\bibnamefont {Gasparinetti}}, \bibinfo {author}
  {\bibfnamefont {M.}~\bibnamefont {Pechal}}, \bibinfo {author} {\bibfnamefont
  {J.}~\bibnamefont {Heinsoo}}, \bibinfo {author} {\bibfnamefont
  {S.}~\bibnamefont {Storz}}, \bibinfo {author} {\bibfnamefont
  {A.}~\bibnamefont {Blais}},\ and\ \bibinfo {author} {\bibfnamefont
  {A.}~\bibnamefont {Wallraff}},\ }\bibfield  {title} {\bibinfo {title} {Fast
  and unconditional all-microwave reset of a superconducting qubit},\ }\href
  {https://doi.org/10.1103/PhysRevLett.121.060502} {\bibfield  {journal}
  {\bibinfo  {journal} {Phys. Rev. Lett.}\ }\textbf {\bibinfo {volume} {121}},\
  \bibinfo {pages} {060502} (\bibinfo {year} {2018})}\BibitemShut {NoStop}%
\bibitem [{\citenamefont {Krinner}\ \emph {et~al.}(2019)\citenamefont
  {Krinner}, \citenamefont {Storz}, \citenamefont {Kurpiers}, \citenamefont
  {Magnard}, \citenamefont {Heinsoo}, \citenamefont {Keller}, \citenamefont
  {L{\"u}tolf}, \citenamefont {Eichler},\ and\ \citenamefont
  {Wallraff}}]{Krinner2019}%
  \BibitemOpen
  \bibfield  {author} {\bibinfo {author} {\bibfnamefont {S.}~\bibnamefont
  {Krinner}}, \bibinfo {author} {\bibfnamefont {S.}~\bibnamefont {Storz}},
  \bibinfo {author} {\bibfnamefont {P.}~\bibnamefont {Kurpiers}}, \bibinfo
  {author} {\bibfnamefont {P.}~\bibnamefont {Magnard}}, \bibinfo {author}
  {\bibfnamefont {J.}~\bibnamefont {Heinsoo}}, \bibinfo {author} {\bibfnamefont
  {R.}~\bibnamefont {Keller}}, \bibinfo {author} {\bibfnamefont
  {J.}~\bibnamefont {L{\"u}tolf}}, \bibinfo {author} {\bibfnamefont
  {C.}~\bibnamefont {Eichler}},\ and\ \bibinfo {author} {\bibfnamefont
  {A.}~\bibnamefont {Wallraff}},\ }\bibfield  {title} {\bibinfo {title}
  {Engineering cryogenic setups for 100-qubit scale superconducting circuit
  systems},\ }\href {https://doi.org/10.1140/epjqt/s40507-019-0072-0}
  {\bibfield  {journal} {\bibinfo  {journal} {EPJ Quantum Technology}\ }\textbf
  {\bibinfo {volume} {6}},\ \bibinfo {pages} {2} (\bibinfo {year}
  {2019})}\BibitemShut {NoStop}%
\bibitem [{\citenamefont {Motzoi}\ \emph {et~al.}(2009)\citenamefont {Motzoi},
  \citenamefont {Gambetta}, \citenamefont {Rebentrost},\ and\ \citenamefont
  {Wilhelm}}]{Motzoi2009}%
  \BibitemOpen
  \bibfield  {author} {\bibinfo {author} {\bibfnamefont {F.}~\bibnamefont
  {Motzoi}}, \bibinfo {author} {\bibfnamefont {J.~M.}\ \bibnamefont
  {Gambetta}}, \bibinfo {author} {\bibfnamefont {P.}~\bibnamefont
  {Rebentrost}},\ and\ \bibinfo {author} {\bibfnamefont {F.~K.}\ \bibnamefont
  {Wilhelm}},\ }\bibfield  {title} {\bibinfo {title} {Simple pulses for
  elimination of leakage in weakly nonlinear qubits},\ }\href
  {https://doi.org/10.1103/PhysRevLett.103.110501} {\bibfield  {journal}
  {\bibinfo  {journal} {Phys. Rev. Lett.}\ }\textbf {\bibinfo {volume} {103}},\
  \bibinfo {eid} {110501} (\bibinfo {year} {2009})}\BibitemShut {NoStop}%
\bibitem [{\citenamefont {Besse}\ \emph {et~al.}(2020)\citenamefont {Besse},
  \citenamefont {Reuer}, \citenamefont {Collodo}, \citenamefont {Wulff},
  \citenamefont {Wernli}, \citenamefont {Copetudo}, \citenamefont {Malz},
  \citenamefont {Magnard}, \citenamefont {Akin}, \citenamefont {Gabureac},
  \citenamefont {Norris}, \citenamefont {Cirac}, \citenamefont {Wallraff},\
  and\ \citenamefont {Eichler}}]{Besse2020a}%
  \BibitemOpen
  \bibfield  {author} {\bibinfo {author} {\bibfnamefont {J.-C.}\ \bibnamefont
  {Besse}}, \bibinfo {author} {\bibfnamefont {K.}~\bibnamefont {Reuer}},
  \bibinfo {author} {\bibfnamefont {M.~C.}\ \bibnamefont {Collodo}}, \bibinfo
  {author} {\bibfnamefont {A.}~\bibnamefont {Wulff}}, \bibinfo {author}
  {\bibfnamefont {L.}~\bibnamefont {Wernli}}, \bibinfo {author} {\bibfnamefont
  {A.}~\bibnamefont {Copetudo}}, \bibinfo {author} {\bibfnamefont
  {D.}~\bibnamefont {Malz}}, \bibinfo {author} {\bibfnamefont {P.}~\bibnamefont
  {Magnard}}, \bibinfo {author} {\bibfnamefont {A.}~\bibnamefont {Akin}},
  \bibinfo {author} {\bibfnamefont {M.}~\bibnamefont {Gabureac}}, \bibinfo
  {author} {\bibfnamefont {G.~J.}\ \bibnamefont {Norris}}, \bibinfo {author}
  {\bibfnamefont {J.~I.}\ \bibnamefont {Cirac}}, \bibinfo {author}
  {\bibfnamefont {A.}~\bibnamefont {Wallraff}},\ and\ \bibinfo {author}
  {\bibfnamefont {C.}~\bibnamefont {Eichler}},\ }\bibfield  {title} {\bibinfo
  {title} {Realizing a deterministic source of multipartite-entangled photonic
  qubits},\ }\href {https://doi.org/10.1038/s41467-020-18635-x} {\bibfield
  {journal} {\bibinfo  {journal} {Nat. Commun.}\ }\textbf {\bibinfo {volume}
  {11}},\ \bibinfo {pages} {4877} (\bibinfo {year} {2020})}\BibitemShut
  {NoStop}%
\bibitem [{\citenamefont {Bultink}\ \emph {et~al.}(2018)\citenamefont
  {Bultink}, \citenamefont {Tarasinski}, \citenamefont {Haandb{\ae}k},
  \citenamefont {Poletto}, \citenamefont {Haider}, \citenamefont {Michalak},
  \citenamefont {Bruno},\ and\ \citenamefont {DiCarlo}}]{Bultink2018}%
  \BibitemOpen
  \bibfield  {author} {\bibinfo {author} {\bibfnamefont {C.~C.}\ \bibnamefont
  {Bultink}}, \bibinfo {author} {\bibfnamefont {B.}~\bibnamefont {Tarasinski}},
  \bibinfo {author} {\bibfnamefont {N.}~\bibnamefont {Haandb{\ae}k}}, \bibinfo
  {author} {\bibfnamefont {S.}~\bibnamefont {Poletto}}, \bibinfo {author}
  {\bibfnamefont {N.}~\bibnamefont {Haider}}, \bibinfo {author} {\bibfnamefont
  {D.~J.}\ \bibnamefont {Michalak}}, \bibinfo {author} {\bibfnamefont
  {A.}~\bibnamefont {Bruno}},\ and\ \bibinfo {author} {\bibfnamefont
  {L.}~\bibnamefont {DiCarlo}},\ }\bibfield  {title} {\bibinfo {title} {General
  method for extracting the quantum efficiency of dispersive qubit readout in
  circuit qed},\ }\href {https://doi.org/10.1063/1.5015954} {\bibfield
  {journal} {\bibinfo  {journal} {Appl. Phys. Lett.}\ }\textbf {\bibinfo
  {volume} {112}},\ \bibinfo {pages} {092601} (\bibinfo {year}
  {2018})}\BibitemShut {NoStop}%
\bibitem [{\citenamefont {Krinner}\ \emph {et~al.}(2022)\citenamefont
  {Krinner}, \citenamefont {Lacroix}, \citenamefont {Remm}, \citenamefont
  {Paolo}, \citenamefont {Genois}, \citenamefont {Leroux}, \citenamefont
  {Hellings}, \citenamefont {Lazar}, \citenamefont {Swiadek}, \citenamefont
  {Herrmann}, \citenamefont {Norris}, \citenamefont {Andersen}, \citenamefont
  {M\"{u}ller}, \citenamefont {Blais}, \citenamefont {Eichler},\ and\
  \citenamefont {Wallraff}}]{Krinner2022}%
  \BibitemOpen
  \bibfield  {author} {\bibinfo {author} {\bibfnamefont {S.}~\bibnamefont
  {Krinner}}, \bibinfo {author} {\bibfnamefont {N.}~\bibnamefont {Lacroix}},
  \bibinfo {author} {\bibfnamefont {A.}~\bibnamefont {Remm}}, \bibinfo {author}
  {\bibfnamefont {A.~D.}\ \bibnamefont {Paolo}}, \bibinfo {author}
  {\bibfnamefont {E.}~\bibnamefont {Genois}}, \bibinfo {author} {\bibfnamefont
  {C.}~\bibnamefont {Leroux}}, \bibinfo {author} {\bibfnamefont
  {C.}~\bibnamefont {Hellings}}, \bibinfo {author} {\bibfnamefont
  {S.}~\bibnamefont {Lazar}}, \bibinfo {author} {\bibfnamefont
  {F.}~\bibnamefont {Swiadek}}, \bibinfo {author} {\bibfnamefont
  {J.}~\bibnamefont {Herrmann}}, \bibinfo {author} {\bibfnamefont {G.~J.}\
  \bibnamefont {Norris}}, \bibinfo {author} {\bibfnamefont {C.~K.}\
  \bibnamefont {Andersen}}, \bibinfo {author} {\bibfnamefont {M.}~\bibnamefont
  {M\"{u}ller}}, \bibinfo {author} {\bibfnamefont {A.}~\bibnamefont {Blais}},
  \bibinfo {author} {\bibfnamefont {C.}~\bibnamefont {Eichler}},\ and\ \bibinfo
  {author} {\bibfnamefont {A.}~\bibnamefont {Wallraff}},\ }\bibfield  {title}
  {\bibinfo {title} {Realizing repeated quantum error correction in a
  distance-three surface code},\ }\href
  {https://doi.org/10.1038/s41586-022-04566-8} {\bibfield  {journal} {\bibinfo
  {journal} {Nature}\ }\textbf {\bibinfo {volume} {605}},\ \bibinfo {pages}
  {669} (\bibinfo {year} {2022})}\BibitemShut {NoStop}%
\bibitem [{\citenamefont {Laurence}\ and\ \citenamefont
  {Chromy}(2010)}]{Laurence2010}%
  \BibitemOpen
  \bibfield  {author} {\bibinfo {author} {\bibfnamefont {T.~A.}\ \bibnamefont
  {Laurence}}\ and\ \bibinfo {author} {\bibfnamefont {B.~A.}\ \bibnamefont
  {Chromy}},\ }\bibfield  {title} {\bibinfo {title} {Efficient maximum
  likelihood estimator fitting of histograms},\ }\href
  {https://doi.org/10.1038/nmeth0510-338} {\bibfield  {journal} {\bibinfo
  {journal} {Nature Methods}\ }\textbf {\bibinfo {volume} {7}},\ \bibinfo
  {pages} {338} (\bibinfo {year} {2010})}\BibitemShut {NoStop}%
\bibitem [{\citenamefont {Walter}\ \emph {et~al.}(2017)\citenamefont {Walter},
  \citenamefont {Kurpiers}, \citenamefont {Gasparinetti}, \citenamefont
  {Magnard}, \citenamefont {Poto\v{c}nik}, \citenamefont {Salath\'e},
  \citenamefont {Pechal}, \citenamefont {Mondal}, \citenamefont {Oppliger},
  \citenamefont {Eichler},\ and\ \citenamefont {Wallraff}}]{Walter2017}%
  \BibitemOpen
  \bibfield  {author} {\bibinfo {author} {\bibfnamefont {T.}~\bibnamefont
  {Walter}}, \bibinfo {author} {\bibfnamefont {P.}~\bibnamefont {Kurpiers}},
  \bibinfo {author} {\bibfnamefont {S.}~\bibnamefont {Gasparinetti}}, \bibinfo
  {author} {\bibfnamefont {P.}~\bibnamefont {Magnard}}, \bibinfo {author}
  {\bibfnamefont {A.}~\bibnamefont {Poto\v{c}nik}}, \bibinfo {author}
  {\bibfnamefont {Y.}~\bibnamefont {Salath\'e}}, \bibinfo {author}
  {\bibfnamefont {M.}~\bibnamefont {Pechal}}, \bibinfo {author} {\bibfnamefont
  {M.}~\bibnamefont {Mondal}}, \bibinfo {author} {\bibfnamefont
  {M.}~\bibnamefont {Oppliger}}, \bibinfo {author} {\bibfnamefont
  {C.}~\bibnamefont {Eichler}},\ and\ \bibinfo {author} {\bibfnamefont
  {A.}~\bibnamefont {Wallraff}},\ }\bibfield  {title} {\bibinfo {title} {Rapid,
  high-fidelity, single-shot dispersive readout of superconducting qubits},\
  }\href {https://doi.org/10.1103/PhysRevApplied.7.054020} {\bibfield
  {journal} {\bibinfo  {journal} {Phys. Rev. Appl.}\ }\textbf {\bibinfo
  {volume} {7}},\ \bibinfo {pages} {054020} (\bibinfo {year}
  {2017})}\BibitemShut {NoStop}%
\bibitem [{\citenamefont {Lienhard}\ \emph {et~al.}(2022)\citenamefont
  {Lienhard}, \citenamefont {Vepsäläinen}, \citenamefont {Govia},
  \citenamefont {Hoffer}, \citenamefont {Qiu}, \citenamefont {Ristè},
  \citenamefont {Ware}, \citenamefont {Kim}, \citenamefont {Winik},
  \citenamefont {Melville}, \citenamefont {Niedzielski}, \citenamefont {Yoder},
  \citenamefont {Ribeill}, \citenamefont {Ohki}, \citenamefont {Krovi},
  \citenamefont {Orlando}, \citenamefont {Gustavsson},\ and\ \citenamefont
  {Oliver}}]{Lienhard2022}%
  \BibitemOpen
  \bibfield  {author} {\bibinfo {author} {\bibfnamefont {B.}~\bibnamefont
  {Lienhard}}, \bibinfo {author} {\bibfnamefont {A.}~\bibnamefont
  {Vepsäläinen}}, \bibinfo {author} {\bibfnamefont {L.~C.~G.}\ \bibnamefont
  {Govia}}, \bibinfo {author} {\bibfnamefont {C.~R.}\ \bibnamefont {Hoffer}},
  \bibinfo {author} {\bibfnamefont {J.~Y.}\ \bibnamefont {Qiu}}, \bibinfo
  {author} {\bibfnamefont {D.}~\bibnamefont {Ristè}}, \bibinfo {author}
  {\bibfnamefont {M.}~\bibnamefont {Ware}}, \bibinfo {author} {\bibfnamefont
  {D.}~\bibnamefont {Kim}}, \bibinfo {author} {\bibfnamefont {R.}~\bibnamefont
  {Winik}}, \bibinfo {author} {\bibfnamefont {A.}~\bibnamefont {Melville}},
  \bibinfo {author} {\bibfnamefont {B.}~\bibnamefont {Niedzielski}}, \bibinfo
  {author} {\bibfnamefont {J.}~\bibnamefont {Yoder}}, \bibinfo {author}
  {\bibfnamefont {G.~J.}\ \bibnamefont {Ribeill}}, \bibinfo {author}
  {\bibfnamefont {T.~A.}\ \bibnamefont {Ohki}}, \bibinfo {author}
  {\bibfnamefont {H.~K.}\ \bibnamefont {Krovi}}, \bibinfo {author}
  {\bibfnamefont {T.~P.}\ \bibnamefont {Orlando}}, \bibinfo {author}
  {\bibfnamefont {S.}~\bibnamefont {Gustavsson}},\ and\ \bibinfo {author}
  {\bibfnamefont {W.~D.}\ \bibnamefont {Oliver}},\ }\bibfield  {title}
  {\bibinfo {title} {Deep-neural-network discrimination of multiplexed
  superconducting-qubit states},\ }\href
  {https://doi.org/10.1103/PhysRevApplied.17.014024} {\bibfield  {journal}
  {\bibinfo  {journal} {Phys. Rev. Applied}\ }\textbf {\bibinfo {volume}
  {17}},\ \bibinfo {pages} {014024} (\bibinfo {year} {2022})}\BibitemShut
  {NoStop}%
\bibitem [{\citenamefont {Huijben}\ \emph {et~al.}(2021)\citenamefont
  {Huijben}, \citenamefont {Kool}, \citenamefont {Paulus},\ and\ \citenamefont
  {van Sloun}}]{huijben_review_2021}%
  \BibitemOpen
  \bibfield  {author} {\bibinfo {author} {\bibfnamefont {I.~A.~M.}\
  \bibnamefont {Huijben}}, \bibinfo {author} {\bibfnamefont {W.}~\bibnamefont
  {Kool}}, \bibinfo {author} {\bibfnamefont {M.~B.}\ \bibnamefont {Paulus}},\
  and\ \bibinfo {author} {\bibfnamefont {R.~J.~G.}\ \bibnamefont {van Sloun}},\
  }\bibfield  {title} {\bibinfo {title} {A {Review} of the {Gumbel}-max {Trick}
  and its {Extensions} for {Discrete} {Stochasticity} in {Machine}
  {Learning}},\ }\href {http://arxiv.org/abs/2110.01515} {\bibfield  {journal}
  {\bibinfo  {journal} {arXiv:2110.01515 [cs, stat]}\ } (\bibinfo {year}
  {2021})}\BibitemShut {NoStop}%
\end{thebibliography}%
